\documentclass[aps,pra,10pt,showpacs,twocolumn,floatfix]{revtex4-1}
\usepackage[T1]{fontenc}
\usepackage{times}
\usepackage{amsfonts,amsmath,amssymb}
\usepackage{graphicx}
\usepackage[pdfstartview=FitH,colorlinks=true,linkcolor=blue,citecolor=blue,urlcolor=blue]{hyperref}
\usepackage{microtype}

\begin{document}

\title{Propagation of a quantum fluid of light in a cavityless nonlinear optical medium: General theory and response to quantum quenches}

\author{Pierre-\'Elie Larr\'e}
\email{pierre.larre@unitn.it}
\author{Iacopo Carusotto}
\email{carusott@science.unitn.it}
\affiliation{INO-CNR BEC Center and Dipartimento di Fisica, Universit\`a di Trento, Via Sommarive 14, 38123 Povo, Italy}

\date{\today}

\begin{abstract}
Making use of a generalized quantum theory of paraxial light propagation where the radiation-axis and the temporal coordinates play exchanged roles, we discuss the potential of bulk nonlinear optical media in cavityless configurations for quantum statistical mechanics studies of the conservative many-body dynamics of a gas of interacting photons. To illustrate the general features of this point of view, we investigate the response of the fluid of light to the quantum quenches in the photon-photon interaction constant experienced at the front and the back faces of a finite slab of weakly nonlinear material. Extending the standard Bogoliubov theory of dilute Bose-Einstein condensates, peculiar features are predicted for the statistical properties of the light emerging from the nonlinear medium.
\end{abstract}

\pacs{42.65.-k, 
      03.70.+k, 
      42.50.Lc, 
      47.37.+q} 

\maketitle

\section{Introduction}
\label{Sec:Introduction}

After several decades during which the study of systems of many interacting particles has focused on matter fluids such as liquid helium, electron gases in solid-state materials, ultracold atom vapors, nuclear-matter fluids, or quark-gluon plasmas in colliders, photon propagation in suitably designed nonlinear optical systems is presently attracting a growing interest as a novel platform to investigate the physics of interacting Bose gases, the so-called \textsl{quantum fluids of light} (see Ref.~\cite{Carusotto2013} for a review). The interactions between the photons constituting the fluid of light are mediated by the Kerr ($\chi^{(3)}$) optical nonlinearity of the underlying medium.

So far, numerous experimental observations have been performed for semiconductor-planar-microcavity geometries, including the demonstration of a Bose-Einstein-like condensation \cite{Kasprzak2006}, of a superfluid flow \cite{Amo2009}, and of the hydrodynamic nucleation of nonlinear excitations such as solitons \cite{Amo2011} and quantized vortices \cite{Nardin2011, Sanvitto2011} in dilute photon gases. In the meanwhile, active investigations have addressed the possibility of realizing systems characterized by very large optical nonlinearities. Correspondingly, the induced photon-photon interactions are expected to introduce strong quantum correlations within the fluid of light, generating in turn new quantum phases.

In this quest \cite{Carusotto2013, Houck2012, Chang2014}, researchers have been faced to (at least) two main difficulties. On the one hand, obtaining strong enough nonlinearities in scalable systems to study the dynamics of a strongly interacting photon gas in a spatially extended system turns out to be a major experimental challenge. On the other hand, the dynamics of a light field in devices based on cavities is intrinsically a driven-dissipative one, which introduces severe complications in the theoretical description of such systems and which is typically very detrimental for the study of purely quantum features.

An alternative platform for studying many-body physics in photon fluids is based on light propagating in a bulk nonlinear (of Kerr type) optical medium. From elementary classical optics, it is well known that the paraxial propagation of a spectrally narrow beam of light can be described within the so-called paraxial and slowly-varying-envelope approximations (see, e.g., Refs.~\cite{Agrawal1995, Raghavan2000, Rosanov2002, Boyd2008}) by a nonlinear wave equation formally analogous to the Gross-Pitaevskii equation for the order parameter of a dilute Bose-Einstein condensate \cite{Pitaevskii2003}. As originally pioneered in Refs.~\cite{Lai1989a, Lai1989b}, this framework naturally translates upon quantization to a many-body quantum nonlinear Schr\"odinger formalism with the roles of the propagation coordinate and time exchanged.

This framework has been used in a number of theoretical works where laser-physics problems have been reformulated in the hydrodynamics language \cite{Mattar1981}, including, e.g., the investigation of superfluid-like behaviors in the flow of a photon fluid \cite{Pomeau1993, Hakim1997, Leboeuf2010, Carusotto2014, Larre2015}, of nonlinear phenomena with light waves \cite{Dekel2007, Khamis2008, Dekel2009, Cohen2013}, and of the so-called acoustic Hawking radiation \cite{Fouxon2010, Fleurov2012, BarAd2013, Vinish2014}. From the experimental point of view, numerous works have been devoted to the study of nonlinear features that may appear in these systems, with a special attention dedicated to their relation to hydrodynamics and superfluidity aspects \cite{Vaupel1996, Wan2007, Jia2007, Wan2010a, Wan2010b, Jia2012, Vocke2015}. A major first step in the very quantum direction of realizing a gas of strongly interacting photons in a propagating geometry has been recently reported using an optically dressed atom gas in the Rydberg-EIT regime as a bulk nonlinear medium: This has allowed for the experimental observation of strongly repulsing photons \cite{Peyronel2012} and, soon after, of two-photon molecular bound states \cite{Firstenberg2013}. These remarkable experimental advances call for a theoretical approach that is able to describe in its full generality the many-body dynamics of strongly interacting photons propagating in a cavityless configuration.

Building atop the pioneering theoretical works \cite{Lai1989a, Lai1989b} and Refs.~\cite{Wright1991, Crosignani1995, Hagelstein1996, Kolobov1999, Matsko2000, Tsang2006}, we report here a fully general quantum field theory of the propagating photon fluid. In this approach, the roles played by the optical-axis coordinate $z$ and the time parameter $t$ are exchanged: Light propagation in the $z$ direction is naturally described in terms of evolution equations while the $t$ direction corresponds to a third spatial dimension in addition to the transverse $x$ and $y$ directions. In a paraxial-propagation configuration, light diffraction provides an effective mass to the photons in the $(x,y)$ plane and chromatic dispersion leads to a---typically different---effective mass in the $t$ direction. As usual, the Kerr nonlinearity of the medium gives rise to photon-photon interactions. In contrast to microcavity configurations where driving and dissipation play a major role in the photon-fluid dynamics \cite{Carusotto2013}, in paraxial-propagation geometries, the quantum fluid of light follows a fully conservative Hamilton dynamics starting from an initial condition determined by the incident light field and its coherence properties. The quantum state of the optical field at the end of the evolution is experimentally accessible \textit{via} a measurement of the statistical properties of the transmitted light emerging from the nonlinear optical medium.

As a most remarkable example of application of this formalism, we then study the transmission of a coherent light across a finite slab of weakly nonlinear medium. In this very simple configuration, the photons experience a pair of sudden jumps of the interaction parameter upon crossing the front and the back faces of the nonlinear medium. As a result of these two \textsl{quantum quenches}, the fluid of light gets excited and its quantum state after the conservative evolution across the nonlinear material can be reconstructed from the statistical properties of the transmitted light. In the weak-nonlinearity regime, the main excitation process consists in the emission of pairs of correlated counterpropagating Bogoliubov phonons, which reflects in peculiar features in the intensity distribution and in the near- and the far-field two-body correlation functions. In its simplicity, this example illustrates the power of the conservative propagation dynamics in view of generating, detecting, and manipulating strongly correlated quantum phases of matter in photon gases, as well as of studying quantum dynamical features of many-body systems \cite{Kinoshita2006, Polkovnikov2011}.

The paper is organized as follows. First of all, in Sec.~\ref{Sec:ClassicalWaveEquation}, we review the classical propagation equation of a paraxial beam of light in a cavityless nonlinear optical medium of Kerr type. On this basis, we present in Sec.~\ref{Sec:QuantumTheory} a general theory that makes it possible to describe the evolution of the quantum optical field for generic values of the nonlinear interaction parameter. In Sec.~\ref{Sec:BogoliubovTheoryOfQuantumFluctuations}, we discuss how the relatively small fluctuations superimposing upon a coherent-light field in the weak-nonlinearity regime can be treated within the framework of the Bogoliubov theory of dilute Bose-Einstein condensates. As an application of this formalism, we investigate in Sec.~\ref{Sec:ResponseToQuantumQuenchesOfTheKerrNonlinearity} the two-body quantum correlations resulting from the propagation of a laser beam across a slab of weakly nonlinear medium and interpret their features in terms of a dynamical Casimir emission of Bogoliubov collective excitations in a temporally modulated quantum fluid of light. Finally, in Sec.~\ref{Sec:ConclusionsAndOutlooks}, we draw our conclusions and give prospects to the present work.

\section{Classical wave equation}
\label{Sec:ClassicalWaveEquation}

We consider the propagation of a laser wave in a dispersive and inhomogeneous Kerr dielectric for which the (frequency-dependent) electric susceptibility reads [one writes down $\mathbf{x}=(\mathbf{x}_{\perp},z)$, with $\mathbf{x}_{\perp}=(x,y)$]
\begin{equation}
\label{Susceptibility}
\chi(\mathbf{x},\omega)=\chi(\omega)+\delta\chi(\mathbf{x})+\chi^{(3)}(\mathbf{x})\,\mathcal{I},
\end{equation}
where the homogeneous contribution $\chi(\omega)$ takes into account the chromatic dispersion of the medium, the linear modulation $\delta\chi(\mathbf{x})$ comes from the existence of spatial inhomogeneities or of an optical confinement, and $\chi^{(3)}(\mathbf{x})\,\mathcal{I}$ is the Kerr nonlinear shift of the susceptibility, proportional to the local electric intensity $\mathcal{I}=|\mathcal{E}|^2$, i.e., to the square modulus of the (complex representation of the) electric field. For simplicity's sake, the dielectric is assumed to be devoid of free charge carriers and nonmagnetic. We finally suppose that the optical field maintains its polarization in the course of its propagation in the medium so that a scalar approach is valid. As explained, e.g., in Ref.~\cite{Rosanov2002}, this is possible for paraxial beam of light and provided $\chi(\mathbf{x},\omega)$ slowly varies on space scales of the order of the optical wavelength. Spin-orbit-coupling effects resulting from significant deviations from the optical axis and significant spatial variations of the electric susceptibility will be subjected to a future publication \cite{LarreFuture}.

Introducing the envelope $\mathcal{E}(\mathbf{x},t)$ of the laser-wave electric field $E(\mathbf{x},t)$ oscillating at the angular frequency $\omega_{0}$ as
\begin{equation}
\label{ElectricField}
E(\mathbf{x},t)=\tfrac{1}{2}\,\mathcal{E}(\mathbf{x},t)\,e^{i(\beta_{0}z-\omega_{0}t)}+\mathrm{c.c.},
\end{equation}
where ``c.c.'' stands for ``complex conjugate,'' and making use of the standard paraxial and slowly-varying-envelope approximations (see, e.g., Refs.~\cite{Agrawal1995, Raghavan2000, Rosanov2002, Boyd2008}), the Maxwell equations supplemented by Eq.~\eqref{Susceptibility} lead to the following classical wave equation for $\mathcal{E}(\mathbf{x},t)$:
\begin{align}
\notag
i\,\frac{\partial\mathcal{E}}{\partial z}=&\left.-\frac{1}{2\,\beta_{0}}\,\nabla_{\perp}^{2}\mathcal{E}+\frac{D_{0}}{2}\,\frac{\partial^{2}\mathcal{E}}{\partial t^{2}}-\frac{i}{v_{0}}\,\frac{\partial\mathcal{E}}{\partial t}\right. \\
\label{WaveEquation}
&\left.+\,U(\mathbf{x})\,\mathcal{E}+g(\mathbf{x})\,|\mathcal{E}|^{2}\,\mathcal{E}.\right.
\end{align}
In this equation, $\nabla_{\perp}=(\partial_{x},\partial_{y})$ denotes the nabla operator in the $\mathbf{x}_{\perp}=(x,y)$ plane and the functions $U(\mathbf{x})$ and $g(\mathbf{x})$ are defined as $[U(\mathbf{x}),g(\mathbf{x})]=-\beta_{0}\,[\delta\chi(\mathbf{x}),\chi^{(3)}(\mathbf{x})]/[2\,(1+\chi_{0})]$, where $\chi_{0}=\chi(\omega_{0})$; the parameters $\beta_{0}=\beta(\omega_{0})$, $v_{0}=v(\omega_{0})$, and $D_{0}=D(\omega_{0})$ are respectively the propagation constant $\beta(\omega)=[1+\chi(\omega)]^{1/2}\,\omega/c$ ($c$ is the vacuum speed of light) of the laser wave in the $z>0$ direction, the group velocity $v(\omega)=[d\beta(\omega)/d\omega]^{-1}$ of the photons in the medium, and the group-velocity dispersion $D(\omega)=d^{2}\beta(\omega)/d\omega^{2}$ evaluated at the carrier's angular frequency $\omega_{0}$.

The hydrodynamic interpretation of the propagation equation \eqref{WaveEquation} is mostly well known in the limiting case of a purely monochromatic wave at $\omega_{0}$ \cite{Pomeau1993, Hakim1997, Dekel2007, Khamis2008, Dekel2009, Leboeuf2010, Cohen2013, Carusotto2014, Larre2015} and has offered a transparent physical interpretation to a number of nonlinear-optics experiments \cite{Vaupel1996, Wan2007, Jia2007, Wan2010a, Wan2010b, Jia2012, Elazar2012, Elazar2013, Vocke2015}. In this time-independent case, the first and second derivatives of the envelope $\mathcal{E}$ with respect to $t$ vanish, in such a way that the propagation of the optical field decribed by Eq.~\eqref{WaveEquation} recovers the mean-field dynamics of the order parameter of a dilute two-dimensional Bose-Einstein condensate, $U(\mathbf{x})\propto\delta\chi(\mathbf{x})$ playing the role of some external potential and $g(\mathbf{x})\propto\chi^{(3)}(\mathbf{x})$ corresponding to the effective two-dimensional boson-boson interaction constant.

On the other hand, only a very few studies so far \cite{Lai1989a, Lai1989b} have investigated the consequences of this analogy in time-dependent regimes, where, by extension, the propagation of the photon fluid in the positive-$z$ direction can be written in terms of a time evolution in a three-dimensional $(\mathbf{x}_{\perp},t)=(x,y,t)$ space where the physical time parameter $t$ plays the role of a third spatial coordinate in addition to the transverse variables $x$ and $y$. In the following, a special attention will be paid to the new hydrodynamic features that originate from this $t$ dependence.

In order to facilitate $z$ ($t$) to be viewed as a true time (space) parameter, we introduce the following coordinates:
\begin{equation}
\label{NewCoordinates}
\tau(z)=\frac{z}{v_{0}}\quad\text{and}\quad\zeta(t)=v_{0}\,t,
\end{equation}
respectively homogeneous to a time and a length. In these new variables, the paraxial-propagation equation \eqref{WaveEquation} takes the form of a time-dependent Gross-Pitaevskii equation:
\begin{align}
\notag
\frac{i}{v_{0}}\,\frac{\partial\mathcal{E}}{\partial\tau}=&\left.-\frac{1}{2\,\beta_{0}}\,\nabla_{\perp}^{2}\mathcal{E}+\frac{v_{0}^{2}\,D_{0}^{\phantom{2}}}{2}\,\frac{\partial^{2}\mathcal{E}}{\partial\zeta^{2}}-i\,\frac{\partial\mathcal{E}}{\partial\zeta}\right. \\
\label{NewWaveEquation}
&\left.+\,U(\mathbf{x}_{\perp},\tau)\,\mathcal{E}+g(\mathbf{x}_{\perp},\tau)\,|\mathcal{E}|^{2}\,\mathcal{E},\right.
\end{align}
where the electric-field envelope $\mathcal{E}$ now has to be considered as a function of $(\mathbf{r},\tau)=(\mathbf{x}_{\perp},\zeta,\tau)$ and $U$ and $g$ as functions of $(\mathbf{x}_{\perp},\tau)$. This $\tau$ dependence of $U$ and $g$ corresponds in our language to a temporal dependence, which, as we shall see in the following, opens the way to the study of quantum-quench physics in the framework of paraxial optics. On the other hand, the fact that the medium properties do not depend on the physical time $t$ implies that $U$ and $g$ are independent on the spatial $\zeta=v_{0}\,t$ coordinate.

In the Gross-Pitaevskii-like equation \eqref{NewWaveEquation}, the rigid-drift (in the $\zeta$ direction) term $-i\,\partial_{\zeta}\mathcal{E}(\mathbf{r},\tau)$ originates from the group velocity of the photons in the medium and the kinetic operator
\begin{align}
\notag
-\frac{1}{2\,\beta_{0}}\,\nabla_{\perp}^{2}&+\frac{v_{0}^{2}\,D_{0}^{\phantom{2}}}{2}\,\frac{\partial^{2}}{\partial\zeta^{2}} \\
\label{KineticOperator}
&=-
\begin{bmatrix}
\displaystyle{\frac{\partial}{\partial x}} &
\displaystyle{\frac{\partial}{\partial y}} &
\displaystyle{\frac{\partial}{\partial\zeta}}
\end{bmatrix}
\frac{1}{2\,\mathcal{M}}
\begin{bmatrix}
\partial/\partial x \vspace{1mm} \\
\partial/\partial y \vspace{1mm} \\
\partial/\partial\zeta
\end{bmatrix}
\end{align}
involves a contribution in the (actual-time) $\zeta$ direction in addition to the usual one in the $\mathbf{x}_{\perp}=(x,y)$ plane. The anisotropy of the ``mass'' tensor
\begin{equation}
\label{MassTensor}
\mathcal{M}=
\begin{bmatrix}
\mathcal{M}_{x,x} & 0 & 0 \\
0 & \mathcal{M}_{y,y} & 0 \\
0 & 0 & \mathcal{M}_{\zeta,\zeta}
\end{bmatrix}
=
\begin{bmatrix}
\beta_{0} & 0 & 0 \\
0 & \beta_{0} & 0 \\
0 & 0 & -\frac{1}{v_{0}^{2}\,D_{0}^{\phantom{2}}}
\end{bmatrix}
\end{equation}
in Eq.~\eqref{KineticOperator} is a natural consequence of the different origins of the effective masses $\mathcal{M}_{x,x}$, $\mathcal{M}_{y,y}$ in the (spatial) $x$, $y$ directions and $\mathcal{M}_{\zeta,\zeta}$ in the (temporal) $\zeta=v_{0}\,t$ one: The former are due to diffraction while the latter originates from dispersion; note that in vacuum, $D_{0}=0$ and so $\mathcal{M}_{\zeta,\zeta}$ is infinite, by definition.

When the medium is characterized by an anomalous group-velocity dispersion, that is, when $D_{0}<0$, one has $\mathcal{M}_{\zeta,\zeta}>0$ and so the mass matrix $\mathcal{M}$ is positive in all the directions. In the following (see Sec.~\ref{Sec:BogoliubovTheoryOfQuantumFluctuations} for a detailed discussion), we shall see that the dynamical stability of the fluid of light requires such a negative $D_{0}$, but also repulsive photon-photon interactions, which sets $g>0$ and therefore $\chi^{(3)}<0$, that is, that the Kerr nonlinearity is self-defocusing.

\section{Quantum theory}
\label{Sec:QuantumTheory}

In order to describe the quantum features of a light beam propagating in the nonlinear medium considered in Sec.~\ref{Sec:ClassicalWaveEquation}, the classical photon field $\mathcal{E}(\mathbf{r},\tau)$ verifying the paraxial wave equation \eqref{NewWaveEquation} must be replaced with a quantum field operator satisfying suitable boson commutation relations. To this purpose, in the present work, we perform a canonical quantization \cite{Dirac1930, Weinberg1995} of the classical field theory from which Eq.~\eqref{NewWaveEquation} may be derived. The procedure first consists in rewriting the evolution equation \eqref{NewWaveEquation} in Lagrangian and then Hamiltonian form (Sec.~\ref{SubSec:HamiltonianFormulation}). Section \ref{SubSec:NormalizationConstant} is dedicated to the precise determination of the global multiplicative constant appearing in the Lagrangian and the Hamiltonian of the paraxial-propagation problem; while it plays no role at the classical level, it starts having a crucial importance upon quantization. The quantization is finally accomplished in Sec.~\ref{SubSec:CanonicalQuantization} by replacing the conjugate fields of the classical Hamiltonian theory with quantum field operators obeying equal-$\tau$ bosonic commutation relations, standardly deduced from the canonical Poisson-bracket relations.

Of course, similar quantized wave equations describing paraxial light propagation in nonlinear optical media have been considered in the past (see, e.g., Refs.~\cite{Lai1989a, Lai1989b, Wright1991, Matsko2000}) and applied to concrete problems related to quantum soliton propagation \cite{Crosignani1995, Hagelstein1996}: As a crucial addition to these works, we do not restrict our attention to one-dimensional fiber geometries for which only the $\zeta$ coordinate matters \cite{LarreFutureBis}, but fully take into account the dynamics of the optical field in the transverse $(x,y)$ plane. Furthermore, an explicit expression for the normalization constant appearing in the boson commutators is provided. In contrast to works on the quantization of paraxial electromagnetic fields \cite{Deutsch1991, Aiello2005}, our approach is able to naturally describe spatiotemporal light propagation and to include the possible existence of spatial inhomogeneities and/or of an optical confinement as well as of an optical nonlinearity. As we shall see later, all these features will be crucial for the theory to be applicable to the problem considered in Sec.~\ref{Sec:ResponseToQuantumQuenchesOfTheKerrNonlinearity}, which requires an accurate description of the peculiar spatial, angular, and spectral correlations displayed by the laser beam emerging from the back face of the nonlinear medium.

At this stage, the reader who is just interested in knowing the basic elements structuring the quantum theory in order to apply the latter to concrete quantum-optics problems can skip Secs.~\ref{SubSec:HamiltonianFormulation} and \ref{SubSec:NormalizationConstant} and go directly to Sec.~\ref{SubSec:CanonicalQuantization}, where one finds the commutation relations of the quantum field operator associated to the classical photon field $\mathcal{E}(\mathbf{r},\tau)$ [Eqs.~\eqref{CommutationRelations}], the many-body quantum Hamiltonian of the paraxial-propagation problem [Eq.~\eqref{QuantizedHamiltonian}], and the corresponding Heisenberg evolution equation [Eq.~\eqref{HeisenbergEquationMotion}].

\subsection{Hamiltonian formulation}
\label{SubSec:HamiltonianFormulation}

Let us introduce the Lagrangian
\begin{equation}
\label{Lagrangian}
L[\mathcal{E}^{\ast},\partial_{\tau}\mathcal{E}^{\ast},\mathcal{E},\partial_{\tau}\mathcal{E};\tau]=\int d\mathbf{r}\;\mathcal{L}(\mathbf{r},\tau),
\end{equation}
with the Lagrangian density
\begin{align}
\notag
\mathcal{L}=&\left.\mathcal{N}\,\bigg[\frac{1}{v_{0}}\,\mathrm{Im}\bigg(\frac{\partial\mathcal{E}^{\ast}}{\partial\tau}\,\mathcal{E}\bigg)-\frac{1}{2\,\beta_{0}}\,|\nabla_{\perp}\mathcal{E}|^{2}+\frac{v_{0}^{2}\,D_{0}^{\phantom{2}}}{2}\,\bigg|\frac{\partial\mathcal{E}}{\partial\zeta}\bigg|^{2}\right. \\
\label{LagrangianDensity}
&\left.+\,\mathrm{Im}\bigg(\frac{\partial\mathcal{E}^{\ast}}{\partial\zeta}\,\mathcal{E}\bigg)-U(\mathbf{x}_{\perp},\tau)\,|\mathcal{E}|^{2}-\frac{g(\mathbf{x}_{\perp},\tau)}{2}\,|\mathcal{E}|^{4}\bigg].\right.
\end{align}
It is immediate to notice that the Euler-Lagrange equations of motion $\delta L(\tau)/\delta\mathcal{E}^{\ast}(\mathbf{r},\tau)=0$ and $\delta L(\tau)/\delta\mathcal{E}(\mathbf{r},\tau)=0$ deduced from Eqs.~\eqref{Lagrangian} and \eqref{LagrangianDensity} coincide with the evolution equation \eqref{NewWaveEquation} and its complex conjugate, respectively. The global normalization factor $\mathcal{N}$ in the definition \eqref{LagrangianDensity} of the Lagrangian density $\mathcal{L}(\mathbf{r},\tau)$ will be rigorously determined in Sec.~\ref{SubSec:NormalizationConstant} on the basis of microscopic calculations. It plays obviously no role at the classical level but is on the contrary crucial to derive the exact commutation relations of the quantum field operators.

Equation \eqref{NewWaveEquation} being a first-order differential equation in $\tau$, the data of the electric-field envelope $\mathcal{E}(\mathbf{r},\tau)$ at a given time $\tau_{0}$ is sufficient to determine the subsequent evolution of the system. As a result, the Lagrangian $L(\tau)$, functional of $\mathcal{E}^{\ast}$, $\mathcal{E}$, and their derivatives with respect to $\tau$, contains an overabundant number of dynamical variables. Thus, before searching for the conjugate momenta and moving on to the Hamiltonian formalism, it is convenient to eliminate the redundant dynamical variables from the Lagrangian. Following Ref.~\cite{CohenTannoudji1989}, one starts by rewriting the Lagrangian density $\mathcal{L}(\mathbf{r},\tau)$ as a function of $\mathcal{E}_{1}(\mathbf{r},\tau)=\mathrm{Re}[\mathcal{E}(\mathbf{r},\tau)]$, $\mathcal{E}_{2}(\mathbf{r},\tau)=\mathrm{Im}[\mathcal{E}(\mathbf{r},\tau)]$, and their spatiotemporal derivatives. In particular, the term involving $\partial_{\tau}\mathcal{E}^{\ast}$ and $\partial_{\tau}\mathcal{E}$ in Eq.~\eqref{LagrangianDensity} reads $(\mathcal{N}/v_{0})\,(\partial_{\tau}\mathcal{E}_{1}\,\mathcal{E}_{2}-\mathcal{E}_{1}\,\partial_{\tau}\mathcal{E}_{2})$. Thus, by adding $(\mathcal{N}/v_{0})\,\frac{d}{d\tau}\int d\mathbf{r}~\mathcal{E}_{1}\,\mathcal{E}_{2}$ to $L(\tau)$, we get a Lagrangian $L'(\tau)$ which does not depend on $\partial_{\tau}\mathcal{E}_{2}$ and which is by definition equivalent to $L(\tau)$. Its density $\mathcal{L}'(\mathbf{r},\tau)$ reads
\begin{align}
\notag
\mathcal{L}'=&\left.\mathcal{N}\,\Bigg\{\frac{2}{v_{0}}\,\frac{\partial\mathcal{E}_{1}}{\partial\tau}\,\mathcal{E}_{2}-\frac{1}{2\,\beta_{0}}\,(|\nabla_{\perp}\mathcal{E}_{1}|^{2}+|\nabla_{\perp}\mathcal{E}_{2}|^{2})\right. \\
\notag
&\left.+\,\frac{v_{0}^{2}\,D_{0}^{\phantom{2}}}{2}\left[\left(\frac{\partial\mathcal{E}_{1}}{\partial\zeta}\right)^{2}+\left(\frac{\partial\mathcal{E}_{2}}{\partial\zeta}\right)^{2}\right]+\frac{\partial\mathcal{E}_{1}}{\partial\zeta}\,\mathcal{E}_{2}-\mathcal{E}_{1}\,\frac{\partial\mathcal{E}_{2}}{\partial\zeta}\right. \\
\label{NewLagrangianDensity}
&\left.-\,U(\mathbf{x}_{\perp},\tau)\,(\mathcal{E}_{1}^{2}+\mathcal{E}_{2}^{2})-\frac{g(\mathbf{x}_{\perp},\tau)}{2}\,(\mathcal{E}_{1}^{2}+\mathcal{E}_{2}^{2})^{2}\Bigg\}.\right.
\end{align}
By means of the Euler-Lagrange equation relating to $\mathcal{E}_{2}(\mathbf{r},\tau)$, that is, $\delta L'(\tau)/\delta\mathcal{E}_{2}(\mathbf{r},\tau)=0$, one may express $\mathcal{E}_{2}(\mathbf{r},\tau)$ as a function of $\mathcal{E}_{1}(\mathbf{r},\tau)$ and its time derivative. Inserting this expression of $\mathcal{E}_{2}(\mathbf{r},\tau)$ into $L'(\tau)$, we finally obtain a Lagrangian $L''(\tau)$ which only involves the dynamical variable $\mathcal{E}_{1}$ and its time derivative $\partial_{\tau}\mathcal{E}_{1}$.

Now let us denote by $\Pi(\mathbf{r},\tau)$ the conjugate momentum of $\mathcal{E}_{1}(\mathbf{r},\tau)$. By definition,
\begin{equation}
\label{ConjugateMomentumDefinition}
\Pi(\mathbf{r},\tau)=\frac{\delta L''[\mathcal{E}_{1},\partial_{\tau}\mathcal{E}_{1};\tau]}{\delta[\partial_{\tau}\mathcal{E}_{1}(\mathbf{r},\tau)]}.
\end{equation}
Along the extremal ``path,'' $\delta L'(\tau)/\delta\mathcal{E}_{2}(\mathbf{r},\tau)=0$, and as a consequence, $\delta L''(\tau)/\delta[\partial_{\tau}\mathcal{E}_{1}(\mathbf{r},\tau)]=\delta L'(\tau)/\delta[\partial_{\tau}\mathcal{E}_{1}(\mathbf{r},\tau)]$. According to Eq.~\eqref{NewLagrangianDensity}, this yields
\begin{equation}
\label{ConjugateMomentum}
\Pi(\mathbf{r},\tau)=\frac{2\,\mathcal{N}}{v_{0}}\,\mathcal{E}_{2}(\mathbf{r},\tau).
\end{equation}
As conjugated fields, $\mathcal{E}_{1}(\mathbf{r},\tau)$ and $\Pi(\mathbf{r},\tau)$ obey the canonical relations
\begin{subequations}
\label{PoissonBracketRelations}
\begin{align}
\label{PoissonBracketRelations1}
&\left.\{\mathcal{E}_{1}(\mathbf{r},\tau),\Pi(\mathbf{r}',\tau)\}_{\tau}=\delta(\mathbf{r}-\mathbf{r}'),\right. \\
\label{PoissonBracketRelations2}
&\left.\{\mathcal{E}_{1}(\mathbf{r},\tau),\mathcal{E}_{1}(\mathbf{r}',\tau)\}_{\tau}=0,\right. \\
\label{PoissonBracketRelations3}
\text{and}\quad&\left.\{\Pi(\mathbf{r},\tau),\Pi(\mathbf{r}',\tau)\}_{\tau}=0,\right.
\end{align}
\end{subequations}
where $\{\cdot,\cdot\}_{\tau}$ is the equal-$\tau$ Poisson bracket.

We are now able to write the Hamiltonian of the system,
\begin{equation}
\label{Hamiltonian}
H(\tau)=\int d\mathbf{r}\;\mathcal{H}(\mathbf{r},\tau).
\end{equation}
In Eq.~\eqref{Hamiltonian}, the Hamiltonian \cite{NoteHamiltonian} density $\mathcal{H}(\mathbf{r},\tau)$ is by definition obtained by Legendre transforming the Lagrangian density $\mathcal{L}''(\mathbf{r},\tau)=\mathcal{L}'(\mathbf{r},\tau)$, $\mathcal{H}(\mathbf{r},\tau)=\Pi(\mathbf{r},\tau)\,\partial_{\tau}\mathcal{E}_{1}(\mathbf{r},\tau)-\mathcal{L}'(\mathbf{r},\tau)$, which, in the light of Eqs.~\eqref{NewLagrangianDensity} and \eqref{ConjugateMomentum} and recalling that $\mathcal{E}(\mathbf{r},\tau)=\mathcal{E}_{1}(\mathbf{r},\tau)+i\,\mathcal{E}_{2}(\mathbf{r},\tau)$, leads to
\begin{align}
\notag
\mathcal{H}=&\left.\mathcal{N}\,\bigg[\frac{1}{2\,\beta_{0}}\,|\nabla_{\perp}\mathcal{E}|^{2}-\frac{v_{0}^{2}\,D_{0}^{\phantom{2}}}{2}\,\bigg|\frac{\partial\mathcal{E}}{\partial\zeta}\bigg|^{2}-\mathrm{Im}\bigg(\frac{\partial\mathcal{E}^{\ast}}{\partial\zeta}\,\mathcal{E}\bigg)\right. \\
\label{HamiltonianDensity}
&\left.+\,U(\mathbf{x}_{\perp},\tau)\,|\mathcal{E}|^{2}+\frac{g(\mathbf{x}_{\perp},\tau)}{2}\,|\mathcal{E}|^{4}\bigg].\right.
\end{align}

\subsection{Normalization constant}
\label{SubSec:NormalizationConstant}

Before moving on to the canonical quantization of the classical field theory (\ref{LagrangianDensity}, \ref{HamiltonianDensity}), let us determine the multiplicative constant $\mathcal{N}$ in terms of the optical parameters of the electromagnetic wave. $\mathcal{N}$ was introduced as a global factor in Eq.~\eqref{LagrangianDensity} to ensure that $\mathcal{L}(\mathbf{r},\tau)$ has the dimension of an energy per unit volume, as it has to be according to Eq.~\eqref{Lagrangian}. While such a normalization constant plays no role at the classical level, since it cancels out in the Euler-Lagrange equations of motion, it becomes important upon quantization as it determines the actual spacing between the energy levels of the system. The role of the quantized action in the Bohr-Sommerfeld quantization rules is the most well-known examples of such a dependence.

A possible strategy to fix $\mathcal{N}$ is to determine the total energy $\tilde{H}_{\mathrm{tot}}(t)$ carried by the laser wave of complex electric field $\tilde{E}(\mathbf{x},t)=\mathcal{E}(\mathbf{x},t)\,e^{i(\beta_{0}z-\omega_{0}t)}$ [$E(\mathbf{x},t)=\frac{1}{2}\,\tilde{E}(\mathbf{x},t)+\mathrm{c.c.}$; see Eq.~\eqref{ElectricField}] in the simple case where $U(\mathbf{x})=0$ and $g(\mathbf{x})=0$, and compare the latter with the well-known formula of classical electrodynamics for the time-averaged energy of a quasimonochromatic electromagnetic field propagating through a dispersive, homogeneous, linear, nonmagnetic medium, that is (see, e.g., Ref.~\cite{Landau1960}),
\begin{equation}
\label{FieldEnergyDispersiveMedia}
\tilde{H}_{\mathrm{field}}(t)=\frac{1}{4}\int d\mathbf{x}\;\bigg\{\frac{d[\omega\,\varepsilon(\omega)]}{d\omega}(\omega_{0})\,|\tilde{E}|^{2}+\frac{|\tilde{B}|^{2}}{\mu_{0}}\bigg\}.
\end{equation}
In Eq.~\eqref{FieldEnergyDispersiveMedia}, $\varepsilon(\omega)=\varepsilon_{0}\,[1+\chi(\omega)]$ denotes the (frequency-dependent) permittivity of the medium, $\varepsilon_{0}$ being the one of free space, $\tilde{B}(\mathbf{x},t)=\mathcal{B}(\mathbf{x},t)\,e^{i(\beta_{0}z-\omega_{0}t)}$, where $\mathcal{B}(\mathbf{x},t)=\pm\,\beta_{0}\,\mathcal{E}(\mathbf{x},t)/\omega_{0}$ is the slowly varying envelope of the magnetic field (deduced, e.g., from Maxwell-Faraday's law), and $\mu_{0}=1/(c^{2}\,\varepsilon_{0})$ is the vacuum permeability.

In order to get an expression for the physical energy $\tilde{H}_{\mathrm{tot}}(t)$ in our Lagrangian formalism, one has to perform a Legendre transformation of the Lagrangian with respect to the actual time coordinate, $t$, instead of the ``propagation one,'' $\tau$. As the Jacobian determinant $|\partial(\zeta,\tau)/\partial(z,t)|=1$ and since the action $S[\mathcal{E}^{\ast},\mathcal{E}]=\int d\tau~L(\tau)=\int d\mathbf{r}\,d\tau~\mathcal{L}(\mathbf{r},\tau)$ of the optical system can be alternatively defined in the $(\mathbf{x},t)$ coordinates as $S[\mathcal{E}^{\ast},\mathcal{E}]=\int dt~\tilde{L}(t)=\int d\mathbf{x}\,dt~\tilde{\mathcal{L}}(\mathbf{x},t)$, the Lagrangian density \eqref{LagrangianDensity} stays invariant under the coordinate transformations \eqref{NewCoordinates}: $\tilde{\mathcal{L}}(\mathbf{x},t)=\mathcal{L}(\mathbf{r},\tau)$. Thus, one finds from Eq.~\eqref{LagrangianDensity} that the conjugate momentum $\tilde{\Pi}(\mathbf{x},t)$ [$\tilde{\Pi}^{\ast}(\mathbf{x},t)$] of $\mathcal{E}(\mathbf{x},t)$ [$\mathcal{E}^{\ast}(\mathbf{x},t)$] in the $(\mathbf{x},t)$ coordinates is given by
\begin{subequations}
\label{ConjugateMomentumTilde}
\begin{align}
\label{ConjugateMomentumTilde1}
\tilde{\Pi}(\mathbf{x},t)&=\frac{\delta\tilde{L}[\mathcal{E}^{\ast},\partial_{t}\mathcal{E}^{\ast},\mathcal{E},\partial_{t}\mathcal{E};t]}{\delta[\partial_{t}\mathcal{E}(\mathbf{x},t)]} \\
\label{ConjugateMomentumTilde2}
&=\mathcal{N}\,\bigg[\frac{i}{2\,v_{0}}\,\mathcal{E}^{\ast}(\mathbf{x},t)+\frac{D_{0}}{2}\,\frac{\partial\mathcal{E}^{\ast}}{\partial t}(\mathbf{x},t)\bigg],
\end{align}
\end{subequations}
as a consequence of which the Hamiltonian density $\tilde{\mathcal{H}}(\mathbf{x},t)=2\,\mathrm{Re}[\tilde{\Pi}(\mathbf{x},t)\,\partial_{t}\mathcal{E}(\mathbf{x},t)]-\tilde{\mathcal{L}}(\mathbf{x},t)$ associated to $\tilde{\mathcal{L}}(\mathbf{x},t)$ in the $(\mathbf{x},t)$ coordinates reads---when $U(\mathbf{x})=0$ and $g(\mathbf{x})=0$---as
\begin{equation}
\label{HamiltonianDensityTilde}
\tilde{\mathcal{H}}=\mathcal{N}\,\bigg[\frac{1}{2\,\beta_{0}}\,|\nabla_{\perp}\mathcal{E}|^{2}+\frac{D_{0}}{2}\,\bigg|\frac{\partial\mathcal{E}}{\partial t}\bigg|^{2}+\mathrm{Im}\bigg(\mathcal{E}^{\ast}\,\frac{\partial\mathcal{E}}{\partial z}\bigg)\bigg].
\end{equation}
Making the substitution $\mathcal{E}(\mathbf{x},t)=\tilde{E}(\mathbf{x},t)\,e^{-i(\beta_{0}z-\omega_{0}t)}$ into Eq.~\eqref{HamiltonianDensityTilde}, one deduces the Hamiltonian $\tilde{H}(t)=\int d\mathbf{x}~\tilde{\mathcal{H}}(\mathbf{x},t)$ as a functional of the total electric field $\tilde{E}(\mathbf{x},t)$:
\begin{align}
\notag
\tilde{H}(t)=&\left.\mathcal{N}\int d\mathbf{x}\;\bigg[\frac{1}{2\,\beta_{0}}\,|\nabla_{\perp}\tilde{E}|^{2}\right. \\
\notag
&\left.+\,\frac{D_{0}}{2}\,\bigg|\frac{\partial\tilde{E}}{\partial t}\bigg|^{2}+\omega_{0}\,D_{0}\,\mathrm{Im}\bigg(\tilde{E}^{\ast}\,\frac{\partial\tilde{E}}{\partial t}\bigg)+\frac{\omega_{0}^{2}\,D_{0}^{\phantom{2}}}{2}\,|\tilde{E}|^{2}\right. \\
\label{HamiltonianTilde}
&\left.+\,\mathrm{Im}\bigg(\tilde{E}^{\ast}\,\frac{\partial\tilde{E}}{\partial z}\bigg)-\beta_{0}\,|\tilde{E}|^{2}\bigg].\right.
\end{align}
The temporal evolution of the envelope $\mathcal{E}(\mathbf{x},t)$ is given by the Hamilton equation of motion $\partial_{t}\mathcal{E}(\mathbf{x},t)=\{\mathcal{E}(\mathbf{x},t),\tilde{H}(t)\}_{t}$, $\{\cdot,\cdot\}_{t}$ being the equal-$t$ Poisson bracket---not to be confused with the equal-$\tau$ Poisson bracket $\{\cdot,\cdot\}_{\tau}$ defined in Sec.~\ref{SubSec:HamiltonianFormulation}. Inserting $\mathcal{E}(\mathbf{x},t)=\tilde{E}(\mathbf{x},t)\,e^{-i(\beta_{0}z-\omega_{0}t)}$ into this evolution equation, one ends up with the following equation for the total field $\tilde{E}(\mathbf{x},t)$:
\begin{subequations}
\label{HamiltonEquationOfMotion}
\begin{align}
\label{HamiltonEquationOfMotion1}
\frac{\partial\tilde{E}}{\partial t}(\mathbf{x},t)&=\{\tilde{E}(\mathbf{x},t),\tilde{H}(t)\}_{t}-i\,\omega_{0}\,\tilde{E}(\mathbf{x},t) \\
\label{HamiltonEquationOfMotion2}
&=\{\tilde{E}(\mathbf{x},t),\tilde{H}_{\mathrm{tot}}(t)\}_{t},
\end{align}
\end{subequations}
where, making use of $\{\mathcal{E}(\mathbf{x},t),\tilde{\Pi}(\mathbf{x}',t)\}_{t}=\delta(\mathbf{x}-\mathbf{x}')$,
\begin{equation}
\label{HamiltonianTotTilde}
\tilde{H}_{\mathrm{tot}}(t)=\tilde{H}(t)-i\,\omega_{0}\int d\mathbf{x}\;e^{-i(\beta_{0}z-\omega_{0}t)}\,\tilde{\Pi}(\mathbf{x},t)\,\tilde{E}(\mathbf{x},t)
\end{equation}
is the Hamiltonian encoding the time evolution of the total electric field $\tilde{E}(\mathbf{x},t)$. By means of Eqs.~\eqref{ConjugateMomentumTilde} and \eqref{HamiltonianTilde}, we can write it in the form
\begin{equation}
\label{HamiltonianTotTildeBis1}
\tilde{H}_{\mathrm{tot}}(t)=\frac{\mathcal{N}\,\omega_{0}}{2\,v_{0}}\int d\mathbf{x}\;|\tilde{E}|^{2}+\delta\tilde{H}_{\mathrm{tot}}(t),
\end{equation}
where
\begin{align}
\notag
\delta\tilde{H}_{\mathrm{tot}}(t)=&\left.\mathcal{N}\int d\mathbf{x}\;\bigg[\frac{1}{2\,\beta_{0}}\,|\nabla_{\perp}\tilde{E}|^{2}\right. \\
\notag
&\left.+\,\frac{D_{0}}{2}\,\bigg|\frac{\partial\tilde{E}}{\partial t}\bigg|^{2}+\frac{\omega_{0}\,D_{0}}{2\,i}\,\tilde{E}^{\ast}\,\frac{\partial\tilde{E}}{\partial t}\right. \\
\label{HamiltonianTotTildeBis2}
&\left.+\,\mathrm{Im}\bigg(\tilde{E}^{\ast}\,\frac{\partial\tilde{E}}{\partial z}\bigg)-\beta_{0}\,|\tilde{E}|^{2}\bigg].\right.
\end{align}

Within the framework of the paraxial and slowly-varying-envelope approximations, one can check that $\delta\tilde{H}_{\mathrm{tot}}(t)$ defined in Eq.~\eqref{HamiltonianTotTildeBis2} gives a negligible contribution to the total energy $\tilde{H}_{\mathrm{tot}}(t)$ of the electromagnetic wave. By comparing Eqs.~\eqref{FieldEnergyDispersiveMedia} and \eqref{HamiltonianTotTildeBis1}, one finally obtains the normalization constant $\mathcal{N}$ in terms of the angular frequency $\omega_{0}$ and the propagation constant $\beta_{0}=(1+\chi_{0})^{1/2}\,\omega_{0}/c$ of the laser in the medium:
\begin{equation}
\label{NormalizationConstant}
\mathcal{N}=\frac{\beta_{0}}{\mu_{0}^{\phantom{2}}\,\omega_{0}^{2}}=\frac{\sqrt{1+\chi_{0}}}{c\,\mu_{0}\,\omega_{0}}.
\end{equation}
Brief discussions on the constant normalizing the Lagrangian and the Hamiltonian in one-dimensional fiber geometries were given in Ref.~\cite{Lai1989a} and, in a bit more detailed way, in Ref.~\cite{Wright1991}. A physical interpretation of expression \eqref{NormalizationConstant} will be given in the next section.

\subsection{Canonical quantization}
\label{SubSec:CanonicalQuantization}

In order to carry out the canonical quantization of the (classical) Hamiltonian theory presented in Sec.~\ref{SubSec:HamiltonianFormulation}, one replaces the conjugated fields $\mathcal{E}_{1}(\mathbf{r},\tau)$ and $\Pi(\mathbf{r},\tau)$ with quantum field operators $\hat{\mathcal{E}}_{1}(\mathbf{r},\tau)$ and $\hat{\Pi}(\mathbf{r},\tau)$ (by choice, in the Heisenberg picture) and the Poisson bracket $\{\cdot,\cdot\}_{\tau}$ with the commutator $[\cdot,\cdot]/(i\,\hbar)$, where $\hbar$ is the reduced Planck constant. On doing so, Eqs.~\eqref{PoissonBracketRelations} become
\begin{subequations}
\label{CanonicalCommutationRelations}
\begin{align}
\label{CanonicalCommutationRelations1}
&\left.[\hat{\mathcal{E}}_{1}(\mathbf{r},\tau),\hat{\Pi}(\mathbf{r}',\tau)]=i\,\hbar\;\delta(\mathbf{r}-\mathbf{r}'),\right. \\
\label{CanonicalCommutationRelations2}
&\left.[\hat{\mathcal{E}}_{1}(\mathbf{r},\tau),\hat{\mathcal{E}}_{1}(\mathbf{r}',\tau)]=0,\right. \\
\label{CanonicalCommutationRelations3}
\text{and}\quad&\left.[\hat{\Pi}(\mathbf{r},\tau),\hat{\Pi}(\mathbf{r}',\tau)]=0,\right.
\end{align}
\end{subequations}
from which and thanks to Eq.~\eqref{ConjugateMomentum} one deduces the following equal-time, that is, equal-$\tau=z/v_{0}$, commutation relations:
\begin{subequations}
\label{CommutationRelations}
\begin{align}
\label{CommutationRelations1}
&\left.[\hat{\mathcal{E}}(\mathbf{r},\tau),\hat{\mathcal{E}}^{\dag}(\mathbf{r}',\tau)]=\frac{\hbar\,v_{0}}{\mathcal{N}}\,\delta(\mathbf{r}-\mathbf{r}')\right. \\
\label{CommutationRelations2}
\text{and}\quad&\left.[\hat{\mathcal{E}}(\mathbf{r},\tau),\hat{\mathcal{E}}(\mathbf{r}',\tau)]=0,\right.
\end{align}
\end{subequations}
$\hat{\mathcal{E}}^{\dag}(\mathbf{r},\tau)$ being the Hermitian conjugate of $\hat{\mathcal{E}}(\mathbf{r},\tau)$: $\hat{\mathcal{E}}^{\dag}(\mathbf{r},\tau)=\hat{\mathcal{E}}_{1}(\mathbf{r},\tau)-i\,\hat{\mathcal{E}}_{2}(\mathbf{r},\tau)$.

It is worth noting that, instead of reducing the overabundant number of dynamical variables in the Lagrangian $L(\tau)$ in order to avoid dealing with fields which are not independent from each other in the transition from the Lagrangian formalism to the Hamiltonian one and in the canonical quantization \textit{via} the Poisson bracket, we could also have implemented the more basic and robust Dirac-Bergmann quantization procedure \cite{Dirac1930, Sundermeyer1982} to pass from the classical theory to the quantum one, as recently used in Ref.~\cite{Vinish2014}. As detailed in footnote \cite{NoteDiracTheory}, Dirac's procedure produces the same equal-time commutation relations for the quantum field operators as the ones deduced from the canonical quantization method adopted in this work, which fully validates our approach.

From the definition $\mathbf{r}=(\mathbf{x}_{\perp},\zeta=v_{0}\,t)$, one notes that the commutation rules \eqref{CommutationRelations} involve quantum field operators at the same ``propagation'' time $\tau=z/v_{0}$ but at different physical times $t$ and $t'\neq t$. Such a writing is based on the coordinate transformation $(\mathbf{x}_{\perp},z,t)\longmapsto[\mathbf{x}_{\perp},\tau(z)=z/v_{0},\zeta(t)=v_{0}\,t]$, which is of course legitimate only in the paraxial and slowly-varying-envelope approximations and if back-scattered waves are assumed not to exist, i.e., if light propagation is assumed to occur only in the positive-$z$ direction. Such a quantization procedure imposing equal-$z$ and different-$t$ (instead of equal-$t$ and different-$z$) canonical commutation relations was pioneered in Refs.~\cite{Lai1989a, Lai1989b} and critically discussed in Ref.~\cite{Matsko2000}. It is also important to insist on the fact that the slow spatiotemporal variation condition that is assumed for the classical electric-field envelope $\mathcal{E}(\mathbf{x},t)$ directly applies to its quantum counterpart $\hat{\mathcal{E}}(\mathbf{r},\tau)$: A discussion of such a condition in a quantum framework is illustrated in Refs.~\cite{Deutsch1991, Aiello2005}.

By inserting the explicit expression \eqref{NormalizationConstant} of the normalization parameter $\mathcal{N}$ into the commutation rule \eqref{CommutationRelations1}, one easily gets that the multiplicative constant in the right-hand side of Eq.~\eqref{CommutationRelations1} reads $(v_{0}/c)\,\hbar\,\omega_{0}/[\varepsilon_{0}\,(1+\chi_{0})^{1/2}]$, yielding a simple interpretation of $\mathcal{N}$ in terms of the energy density of a single-photon wavepacket of total energy $\hbar\,\omega_{0}$.

In the light of Eqs.~\eqref{Hamiltonian} and \eqref{HamiltonianDensity}, the quantized Hamiltonian of the system reads, after considering the normal order,
\begin{align}
\notag
\hat{H}(\tau)=&\left.\mathcal{N}\int d\mathbf{r}\;\bigg[\frac{1}{2\,\beta_{0}}\,\nabla_{\perp}\hat{\mathcal{E}}^{\dag}\cdot\nabla_{\perp}\hat{\mathcal{E}}-\frac{v_{0}^{2}\,D_{0}^{\phantom{2}}}{2}\,\frac{\partial\hat{\mathcal{E}}^{\dag}}{\partial\zeta}\,\frac{\partial\hat{\mathcal{E}}}{\partial\zeta}\right. \\
\notag
&\left.+\,\frac{i}{2}\,\bigg(\frac{\partial\hat{\mathcal{E}}^{\dag}}{\partial\zeta}\,\hat{\mathcal{E}}-\hat{\mathcal{E}}^{\dag}\,\frac{\partial\hat{\mathcal{E}}}{\partial\zeta}\bigg)\right. \\
\label{QuantizedHamiltonian}
&\left.+\,U(\mathbf{x}_{\perp},\tau)\,\hat{\mathcal{E}}^{\dag}\,\hat{\mathcal{E}}+\frac{g(\mathbf{x}_{\perp},\tau)}{2}\,\hat{\mathcal{E}}^{\dag}\,\hat{\mathcal{E}}^{\dag}\,\hat{\mathcal{E}}\,\hat{\mathcal{E}}\bigg].\right.
\end{align}
Equation \eqref{QuantizedHamiltonian} corresponds to the many-body quantum Hamiltonian describing the evolution in time $\tau=z/v_{0}$ of a many-photon laser beam propagating through the bulk inhomogeneous and nonlinear optical medium of electric susceptibility \eqref{Susceptibility}. In the dielectric, the position of a point is referenced by the coordinates $\mathbf{x}_{\perp}=(x,y)$ and $\zeta=v_{0}\,t$. The two first contributions are the kinetic terms in the transverse $(x,y)$ plane and in the $\zeta$ direction with different effective masses (as discussed in Sec.~\ref{Sec:ClassicalWaveEquation}), the second line describes the rigid global drift along the $\zeta$ axis due the group velocity of light in the medium, and the two last terms respectively account for the spatial modulation of the electric susceptibility and for the two-photon interactions mediated by the Kerr nonlinearity of the dielectric.

In the theory of ultracold Bose fluids, contact interactions are usually considered in place of the actual---but much more complicated---two-body interactions. This approximation is very helpful in simplifying the many-body quantum problem and is well accurate as long as the inter-particle distance is much larger than the range of the boson-boson interactions (see Ref.~\cite{Pitaevskii2003}). Here, the assumed local form of the Kerr optical nonlinearity automatically leads to contact-like interactions between the photons of the light beam, that is, no dilutness condition for the photon gas is required to get the four-field interaction term $\propto\int d\mathbf{r}~\hat{\mathcal{E}}^{\dag}\,\hat{\mathcal{E}}^{\dag}\,\hat{\mathcal{E}}\,\hat{\mathcal{E}}$ in Eq.~\eqref{QuantizedHamiltonian}. Since no hypothesis is made on the intensity of the photon-photon interaction parameter $g$, the many-body Hamiltonian \eqref{QuantizedHamiltonian} can describe a quantum fluid of weakly interacting photons, that is, in the Gross-Pitaevskii/Bogoliubov regime, as well as a strongly interacting one. In what follows, we will focus our attention on the weak-nonlinearity regime, in which the quantum fluctuations of the fluid of light can be described within the framework of the well-known Bogoliubov theory of dilute Bose-Einstein condensates (see, e.g., Refs.~\cite{Pitaevskii2003, Castin2001, Fetter2003}). Motivated by the intense experimental investigations that are presently in progress \cite{Peyronel2012, Firstenberg2013}, the strong-interaction regime will be the subject of future works \cite{Lebreuilly, LebreuillyBis}.

For the sake of completeness, it is useful to explicitly write the evolution equation of the operator $\hat{\mathcal{E}}(\mathbf{r},\tau)$. In the Heisenberg picture, it is obtained from the system's Hamiltonian as $i\,\hbar\,\partial_{\tau}\hat{\mathcal{E}}(\mathbf{r},\tau)=-[\hat{H}(\tau),\hat{\mathcal{E}}(\mathbf{r},\tau)]$. Using Eq.~\eqref{QuantizedHamiltonian} and taking advantage of the commutation relations \eqref{CommutationRelations}, this gives
\begin{align}
\notag
\frac{i}{v_{0}}\,\frac{\partial\hat{\mathcal{E}}}{\partial\tau}=&\left.-\frac{1}{2\,\beta_{0}}\,\nabla_{\perp}^{2}\hat{\mathcal{E}}+\frac{v_{0}^{2}\,D_{0}^{\phantom{2}}}{2}\,\frac{\partial^{2}\hat{\mathcal{E}}}{\partial\zeta^{2}}-i\,\frac{\partial\hat{\mathcal{E}}}{\partial\zeta}\right. \\
\label{HeisenbergEquationMotion}
&\left.+\,U(\mathbf{x}_{\perp},\tau)\,\hat{\mathcal{E}}+g(\mathbf{x}_{\perp},\tau)\,\hat{\mathcal{E}}^{\dag}\,\hat{\mathcal{E}}\,\hat{\mathcal{E}}.\right.
\end{align}
Equation \eqref{HeisenbergEquationMotion}, which governs the time evolution of $\hat{\mathcal{E}}(\mathbf{r},\tau)$ in the $\mathbf{r}=(\mathbf{x}_{\perp},\zeta)$ space, is simply the quantized version of the classical equation \eqref{NewWaveEquation}. As originally pointed out in Ref.~\cite{Lai1989a} for a one-dimensional waveguide geometry, it has the form of a quantum nonlinear Schr\"odinger equation.

\section{Bogoliubov theory of quantum fluctuations}
\label{Sec:BogoliubovTheoryOfQuantumFluctuations}

\subsection{General framework}
\label{SubSec:GeneralFramework}

In an illuminated dielectric medium devoid of free charges as the one considered in this paper, quantum noise of the electromagnetic field only arises from the quantum uncertainty of the optical field, that is, in more physical terms, from the discreteness of the photon. In the case of a strongly coherent light beam propagating across a weakly nonlinear three-dimensional bulk medium, quantum noise is typically small and can be described in terms of weak-amplitude quantum fluctuations oscillating on top of a strongly classical wave.

Mutuating well-known results from the theory of weakly interacting ultracold atomic gases \cite{Pitaevskii2003, Castin2001, Fetter2003}, one may develop a Bogoliubov-like theory based on an expansion of the envelope operator $\hat{\mathcal{E}}(\mathbf{r},\tau)$ of the form
\begin{equation}
\label{BogoliubovPrescription}
\hat{\mathcal{E}}(\mathbf{r},\tau)=\mathcal{E}(\mathbf{r},\tau)+\delta\hat{\mathcal{E}}(\mathbf{r},\tau).
\end{equation}
In this expression, the classical field $\mathcal{E}(\mathbf{r},\tau)$, which satisfies the Gross-Pitaevskii-like equation \eqref{NewWaveEquation}, corresponds to the coherent component of the electric-field envelope and $\delta\hat{\mathcal{E}}(\mathbf{r},\tau)$ is a small quantum correction to $\mathcal{E}(\mathbf{r},\tau)$. As the whole quantum nature of the optical field is captured in the fluctuation operator $\delta\hat{\mathcal{E}}(\mathbf{r},\tau)$, the equal-$\tau$ commutation relations \eqref{CommutationRelations} then totally transfer to the latter, giving
\begin{subequations}
\label{CommutationRelationsFluctuation}
\begin{align}
\label{CommutationRelationsFluctuation1}
&\left.[\delta\hat{\mathcal{E}}(\mathbf{r},\tau),\delta\hat{\mathcal{E}}^{\dag}(\mathbf{r}',\tau)]=\frac{\hbar\,v_{0}}{\mathcal{N}}\,\delta(\mathbf{r}-\mathbf{r}')\right. \\
\label{CommutationRelationsFluctuation2}
\text{and}\quad&\left.[\delta\hat{\mathcal{E}}(\mathbf{r},\tau),\delta\hat{\mathcal{E}}(\mathbf{r}',\tau)]=0.\right.
\end{align}
\end{subequations}
Linearizing the Heisenberg equation of motion \eqref{HeisenbergEquationMotion} with respect to $\delta\hat{\mathcal{E}}(\mathbf{r},\tau)$ and its Hermitian conjugate, one readily gets the so-called Bogoliubov-de Gennes equation
\begin{align}
\notag
\frac{i}{v_{0}}\,\frac{\partial\delta\hat{\mathcal{E}}}{\partial\tau}=&\left.-\frac{1}{2\,\beta_{0}}\,\nabla_{\perp}^{2}\delta\hat{\mathcal{E}}+\frac{v_{0}^{2}\,D_{0}^{\phantom{2}}}{2}\,\frac{\partial^{2}\delta\hat{\mathcal{E}}}{\partial\zeta^{2}}-i\,\frac{\partial\delta\hat{\mathcal{E}}}{\partial\zeta}\right. \\
\notag
&\left.+\,U(\mathbf{x}_{\perp},\tau)\,\delta\hat{\mathcal{E}}+2\,g(\mathbf{x}_{\perp},\tau)\,|\mathcal{E}(\mathbf{r},\tau)|^{2}\,\delta\hat{\mathcal{E}}\right. \\
\label{BogoliubovDeGennesEquation}
&\left.+\,g(\mathbf{x}_{\perp},\tau)\,\mathcal{E}^{2}(\mathbf{r},\tau)\,\delta\hat{\mathcal{E}}^{\dag},\right.
\end{align}
which encodes the time evolution of the quantum fluctuation $\delta\hat{\mathcal{E}}(\mathbf{r},\tau)$ in the three-dimensional $\mathbf{r}=(\mathbf{x}_{\perp},\zeta)$ space and is in fact the heart of the Bogoliubov approach. With respect to similar (classical) equations considered in the literature \cite{Carusotto2014, Vinish2014, Vocke2015}, this equation explicitly includes the $\zeta=v_{0}\,t$ dependence of the field and the corresponding effective mass, given by the group-velocity dispersion $D(\omega)$ at $\omega=\omega_{0}$.

In Sec.~\ref{SubSec:SpatiallyHomogeneousSystem}, we review the Bogoliubov theory of linearized fluctuations in position- and time-independent configurations. Although well established and known in the context of matter fluids, it is important to quickly review it within the nonlinear-propagating-geometry context because (i) of the nontrivial role of the effective mass $\mathcal{M}_{\zeta,\zeta}=-1/(v_{0}^{2}\,D_{0}^{\phantom{2}})$ in the dynamics of the luminous fluid in the temporal $\zeta=v_{0}\,t$ direction and (ii) of the existence of on-going experiments which aim at probing the phononic part of the Bogoliubov excitation spectrum of a propagating fluid of light \cite{Vocke2015, Biasi}.

\subsection{Spatially homogeneous system}
\label{SubSec:SpatiallyHomogeneousSystem}

In the simplest case where the classical background field $\mathcal{E}(\mathbf{r},\tau)$ is at some time $\tau=\tau_{0}$ homogeneous in all the ($x$, $y$, and $\zeta$) directions, $\mathcal{E}(\mathbf{r},\tau_{0})=\mathcal{E}\,e^{i\varphi_{0}}$ with $\mathcal{E},\varphi_{0}=\mathrm{const}\in\mathbb{R}$, and when the nonlinear optical medium is spatially homogeneous with a constant Kerr coefficient $\chi^{(3)}$,
\begin{subequations}
\label{UniformSituation}
\begin{align}
\label{UniformSituation1}
&\left.U(\mathbf{x}_{\perp},\tau)=0\right. \\
\label{UniformSituation2}
\text{and}\quad&\left.g(\mathbf{x}_{\perp},\tau)=-\frac{\beta_{0}}{2\,(1+\chi_{0})}\,\chi^{(3)}=g_{0},\right.
\end{align}
\end{subequations}
analytical solutions for the classical wave equation \eqref{NewWaveEquation} and the Bogoliubov-de Gennes equation \eqref{BogoliubovDeGennesEquation} are available (see, e.g., Refs.~\cite{Pitaevskii2003, Castin2001, Fetter2003}).

In such a configuration, the electric-field envelope $\mathcal{E}(\mathbf{r},\tau)$ follows a simple harmonic evolution with a linearly-evolving (in time $\tau$) global phase: $\mathcal{E}(\mathbf{r},\tau)=\mathcal{E}\,e^{i\varphi(\tau)}$, where $\varphi(\tau)=\varphi_{0}-v_{0}\,g_{0}\,\mathcal{E}^{2}\,(\tau-\tau_{0})$. In the theory of dilute Bose-Einstein condensates, the wavenumber $g_{0}\,\mathcal{E}^{2}$ correponds to the chemical potential of the Bose gas \cite{Pitaevskii2003}.

In the homogeneous situation \eqref{UniformSituation}, the elementary excitations of the fluid of light are plane waves of wavevector $\mathbf{q}=(\mathbf{q}_{\perp},q_{\zeta})=(q_{x},q_{y},q_{\zeta})$ in the $\mathbf{r}=(\mathbf{x}_{\perp},\zeta)=(x,y,\zeta)$ space, the solution of the Bogoliubov-de Gennes equation \eqref{BogoliubovDeGennesEquation} obeying the mode expansion
\begin{align}
\notag
\delta\hat{\mathcal{E}}(\mathbf{r},\tau)=&\left.e^{i\varphi(\tau)}\int\frac{d\mathbf{q}}{(2\pi)^{3}}\,\Big[u_{\mathbf{q}}\,e^{i\mathbf{q}\cdot\mathbf{r}}\,\hat{b}(\mathbf{q},\tau)\right. \\
\label{BogoliubovPlaneWaves}
&\left.+\,v_{\mathbf{q}}\,e^{-i\mathbf{q}\cdot\mathbf{r}}\,\hat{b}^{\dag}(\mathbf{q},\tau)\Big].\right.
\end{align}
The mode operators $\hat{b}(\mathbf{q},\tau)$ satisfy the same-$\tau$ commutation relations
\begin{subequations}
\label{CommutationBogoliubovOperators}
\begin{align}
\label{CommutationBogoliubovOperators1}
&\left.[\hat{b}(\mathbf{q},\tau),\hat{b}^{\dag}(\mathbf{q}',\tau)]=(2\pi)^{3}\,\frac{\hbar\,v_{0}}{\mathcal{N}}\,\delta(\mathbf{q}-\mathbf{q}')\right. \\
\label{CommutationBogoliubovOperators2}
\text{and}\quad&\left.[\hat{b}(\mathbf{q},\tau),\hat{b}(\mathbf{q}',\tau)]=0,\right.
\end{align}
\end{subequations}
and evolve harmonically as
\begin{equation}
\label{BogoliubovOperators}
\hat{b}(\mathbf{q},\tau)=\hat{b}(\mathbf{q},\tau_{0})\exp[-i\,\omega_{\mathbf{q}}\,(\tau-\tau_{0})],
\end{equation}
with a frequency $\omega_{\mathbf{q}}=v_{0}\,[K_{\mathrm{B}}(\mathbf{q})+q_{\zeta}]$ determined by the $\mathbf{q}$-dependent wavenumber
\begin{equation}
\label{BogoliubovSpectrum}
K_{\mathrm{B}}(\mathbf{q})=\sqrt{K_{\mathrm{kin}}(\mathbf{q})\,[K_{\mathrm{kin}}(\mathbf{q})+2\,g_{0}\,\mathcal{E}^{2}]},
\end{equation}
where
\begin{equation}
\label{KineticEnergy}
K_{\mathrm{kin}}(\mathbf{q})=\frac{1}{2\,\beta_{0}}\,\mathbf{q}_{\perp}^{2}-\frac{v_{0}^{2}\,D_{0}^{\phantom{2}}}{2}\,q_{\zeta}^{2}.
\end{equation}
The so-called Bogoliubov amplitudes $u_{\mathbf{q}}$ and $v_{\mathbf{q}}$ [$\mathbf{r}$ and $\tau$ independent in the situation \eqref{UniformSituation}] are finally given by
\begin{equation}
\label{BogoliubovAmplitudes}
u_{\mathbf{q}},v_{\mathbf{q}}=\frac{1}{2}\,\frac{K_{\mathrm{kin}}(\mathbf{q})\pm K_{\mathrm{B}}(\mathbf{q})}{\sqrt{K_{\mathrm{kin}}(\mathbf{q})\,K_{\mathrm{B}}(\mathbf{q})}}
\end{equation}
By definition, they satisfy the normalization condition $u_{\mathbf{q}}^{2}-v_{\mathbf{q}}^{2}=1$. Because of the photon-photon interactions ($g_{0}\neq0$), $v_{\mathbf{q}}$ is nonzero: This indicates that the ground state of the Bogoliubov theory differs from the trivial vacuum without particles in the $\mathbf{q}\neq0$ modes \cite{Pitaevskii2003}.

\begin{figure}
\includegraphics[width=\linewidth]{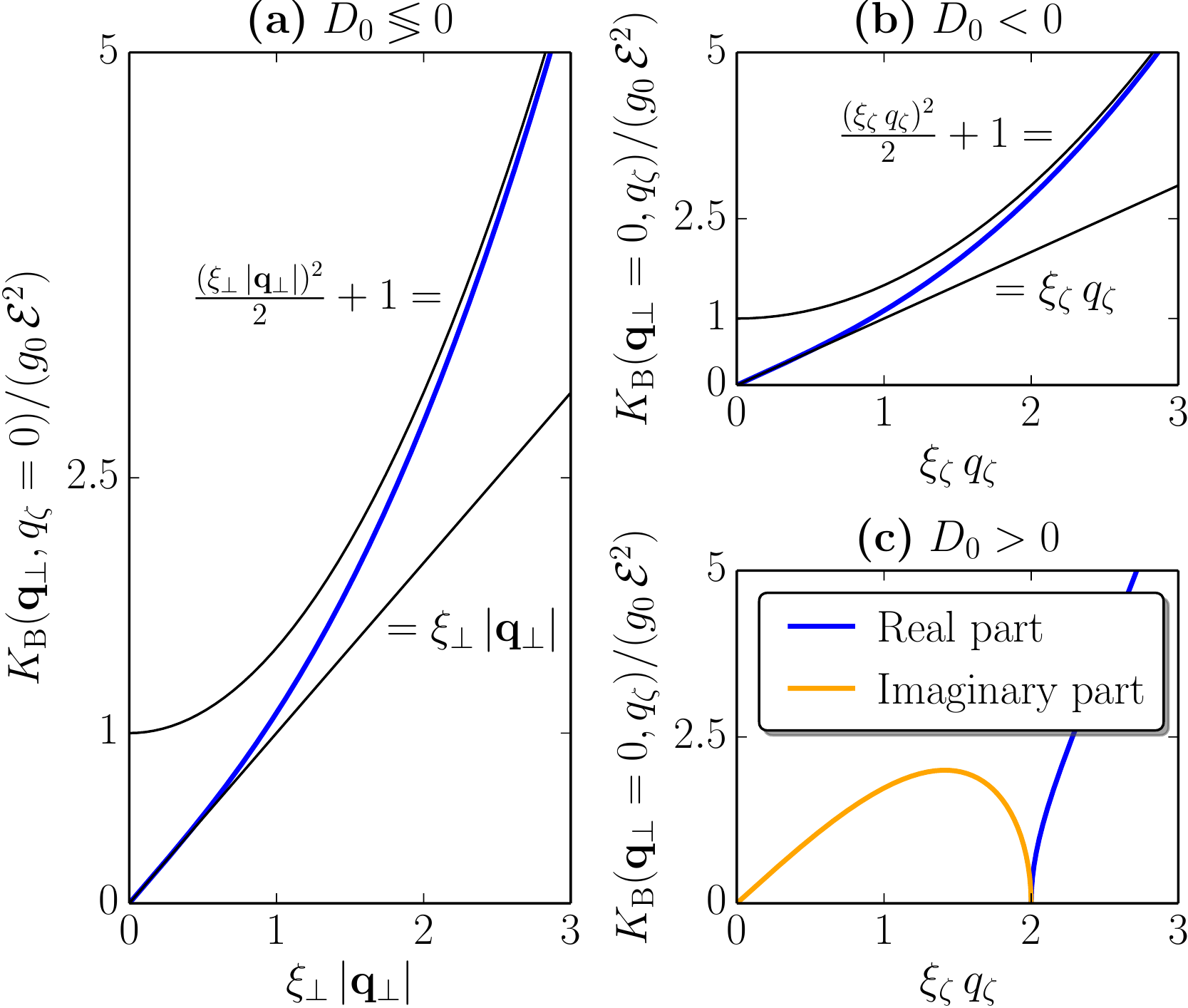}
\caption{(Color online) Dispersion curves $K_{\mathrm{B}}(\mathbf{q})$ (in units of $g_{0}\,\mathcal{E}^{2}$) of the elementary excitations propagating on top of a homogeneous quantum fluid of weakly interacting photons (with $g_{0}>0$) at rest. In the anomalous-dispersion case [$D_{0}<0$, panels (a) and (b)], $K_{\mathrm{B}}(\mathbf{q})$ corresponds to the well-known Bogoliubov law, linear at small $\mathbf{q}$'s (which is indicated by the black straight lines) and quadratic at large $\mathbf{q}$'s (black parabola). In the normal-dispersion case [$D_{0}>0$, panels (a) and (c)], the effective mass $\mathcal{M}_{\zeta,\zeta}$ in the $\zeta$ direction is negative and $K_{\mathrm{B}}(\mathbf{q})$ has an imaginary component [cf.~panel (c)]: In this case, the quantum fluid of light is dynamically unstable.}
\label{BogoliubovLaw}
\end{figure}

Modulo the $q_{\zeta}$ contribution to the oscillation pulsation $\omega_{\mathbf{q}}$, originating from the presence of the drift term $-i\,\partial_{\zeta}\delta\hat{\mathcal{E}}(\mathbf{r},\tau)$ in the Bogoliubov-de Gennes equation \eqref{BogoliubovDeGennesEquation}, the wavenumber $K_{\mathrm{B}}(\mathbf{q})$ defined in Eq.~\eqref{BogoliubovSpectrum} corresponds to the well-known Bogoliubov dispersion relation for the elementary excitations propagating on top of a uniform dilute Bose-Einstein condensate at rest, the quadratic function $K_{\mathrm{kin}}(\mathbf{q})$ given by Eq.~\eqref{KineticEnergy} playing the role of the single-particle kinetic energy of matter-wave superfluids. Note that this result stems from the conservative nature of the considered dynamics and is in contrast to the rich variety of dispersions predicted for driven-dissipative fluids of light in microcavity architectures \cite{Carusotto2013}. The graphical representation of $K_{\mathrm{B}}(\mathbf{q})$ is given in Fig.~\ref{BogoliubovLaw} for repulsive photon-photon interactions, i.e., for $g_0>0$, in the two cases $D_{0}\lessgtr0$.

In the anomalous-dispersion case, that is, when $D_{0}<0$, the photon effective mass $\mathcal{M}_{\zeta,\zeta}=-1/(v_{0}^{2}\,D_{0}^{\phantom{2}})$ in the $\zeta$ direction is---as the photon effective mass $\mathcal{M}_{x,x}=\mathcal{M}_{y,y}=\beta_{0}$ in the transverse $(x,y)$ plane---positive, and the ``kinetic energy'' $K_{\mathrm{kin}}(\mathbf{q})$ stays as a consequence positive for any $\mathbf{q}$. If one also considers a self-defocusing nonlinearity, that is, if $g_{0}>0$, the Bogoliubov dispersion relation $K_{\mathrm{B}}(\mathbf{q})$ never acquires an imaginary part, which means that the photon-photon collision processes mediated by the underlying nonlinear medium do not give rise to unstable behaviors in the photon fluid.

In the case mentioned above where $D_{0}<0$ and $g_{0}>0$, at low wavevectors, that is, when $\xi_{\perp}\,|\mathbf{q}_{\perp}|,\xi_{\zeta}\,|q_{\zeta}|\ll1$, where $\xi_{\perp}=1/(\mathcal{M}_{x,x}\,g_{0}\,\mathcal{E}^{2})^{1/2}=1/(\mathcal{M}_{y,y}\,g_{0}\,\mathcal{E}^{2})^{1/2}$ and $\xi_{\zeta}=1/(\mathcal{M}_{\zeta,\zeta}\,g_{0}\,\mathcal{E}^{2})^{1/2}$ are the healing lengths \cite{Pitaevskii2003} of the photon fluid in the $\mathbf{x}_{\perp}=(x,y)$ plane and in the $\zeta$ direction, respectively, the Bogoliubov dispersion relation $K_{\mathrm{B}}(\mathbf{q})$ is phonon-like in all the directions [as shown by the black straight lines in panels (a) and (b) of Fig.~\ref{BogoliubovLaw}], but with different ``sound velocities'' \cite{Pitaevskii2003} $s_{\perp}=g_{0}\,\mathcal{E}^{2}\,\xi_{\perp}$ and $s_{\zeta}=g_{0}\,\mathcal{E}^{2}\,\xi_{\zeta}\neq s_{\perp}$:
\begin{equation}
\label{SoundLikeDispersion}
K_{\mathrm{B}}(\mathbf{q})\simeq\sqrt{(s_{\perp}\,|\mathbf{q}_{\perp}|)^{2}+(s_{\zeta}\,q_{\zeta})^{2}}.
\end{equation}
A pump-probe measurement of $s_{\perp}$ was recently reported in Ref.~\cite{Vocke2015}. Another experiment aiming at measuring $s_{\zeta}$ in a one-dimensional waveguide configuration is presently in progress \cite{Biasi}. In the opposite limit, that is, when $\xi_{\perp}\,|\mathbf{q}_{\perp}|,\xi_{\zeta}\,|q_{\zeta}|\gg1$, the dispersion $K_{\mathrm{B}}(\mathbf{q})$ of the elementary excitations takes a single-particle-like shape [as illustrated by the black parabola in panels (a) and (b) of Fig.~\ref{BogoliubovLaw}]:
\begin{equation}
\label{ParticleLikeDispersion}
K_{\mathrm{B}}(\mathbf{q})\simeq K_{\mathrm{kin}}(\mathbf{q})+g_{0}\,\mathcal{E}^{2}.
\end{equation}
In the present paraxial-optics context, the Hartree interaction term $g_{0}\,\mathcal{E}^{2}$ corresponds to the usual modification of the propagation constant due to the Kerr nonlinearity of the underlying medium.

The situation is a bit more complicated in the case of a normal dispersion, that is, when $D_{0}>0$: As presented in panel (c) of Fig.~\ref{BogoliubovLaw}, there exists in that case a range of wavevectors $\mathbf{q}$ in the $\zeta$ direction for which the dispersion law $K_{\mathrm{B}}(\mathbf{q})$ is purely imaginary; this signals that the quantum fluid of light is dynamically unstable against the formation of a train of pulses (see, e.g., Refs.~\cite{Agrawal1995, Hasegawa1984, Ganapathy2006, Zakharov2009}).

To complete the picture, we can also mention that for focusing nonlinearities ($g_{0}<0$) the low-wavevector modes in the transverse $(x,y)$ plane are unstable for all $D_{0}$: This gives rise to a modulational instability which, in the optical language, goes under the name of filamentation instability (see, e.g., Refs.~\cite{ShuangChun2001, Berge2003, Couairon2007, Bree2012}). On the other hand, in this case, the instability in the temporal direction only occurs for a anomalous group-velocity dispersion ($D_{0}<0$).

\section{Response to quantum quenches of the Kerr nonlinearity}
\label{Sec:ResponseToQuantumQuenchesOfTheKerrNonlinearity}

As a first example of application of the general quantum formalism developed in Secs.~\ref{Sec:QuantumTheory} and \ref{Sec:BogoliubovTheoryOfQuantumFluctuations}, we investigate in the present section the propagation of a wide laser beam across a slab of Kerr medium immersed in vacuum (see Fig.~\ref{ExperimentalSetup}). Having in mind the reformulation of the propagation of the optical field along the $z$ axis in terms of a temporal evolution in time $\tau=z/v_{0}$, it is straightforward to see how such a configuration constitutes a very simple realization of a pair of quantum quenches of the nonlinear interaction parameter in the system's Hamiltonian \eqref{QuantizedHamiltonian}. As the optical nonlinearity is nonzero only inside the Kerr material, the first (second) quench occurs at the entrance (exit) face of the dielectric, where the value of the photon-photon interaction constant suddenly jumps from $0$ to $\propto\chi^{(3)}\neq0$ (from $\propto\chi^{(3)}\neq0$ to $0$). While the present work focuses on the weak-nonlinearity regime, regime within which the quantum fluctuations of the fluid of light are accurately described by the Bogoliubov theory of linearized fluctuations reviewed in the previous section, application of the general theory \eqref{CommutationRelations}--\eqref{HeisenbergEquationMotion} to the strongly interacting regimes experimentally realized in Refs.~\cite{Peyronel2012, Firstenberg2013} is definitely possible and will be subjected to future works \cite{Lebreuilly, LebreuillyBis}.

Given the extreme simplicity of the proposed setup as compared to corresponding experiments using ultracold atomic vapors \cite{Kinoshita2006}, this study demonstrates the promise of nonlinear optical systems to investigate different features of conservative quantum dynamics, including the response to quenches in the Hamiltonian's parameters and the subsequent thermalization dynamics \cite{Polkovnikov2011}: In contrast to microcavity systems where cavity losses play a crucial role in the dynamics and quickly wash out quantum correlations \cite{Carusotto2013}, in the present propagating geometry, losses can be made arbitrarily small simply by choosing a suitable transparent dielectric medium.

Building on decades of expertise in quantum-optics experiments, we discuss below how a detailed information on the response of the photon fluid to the quenches of the interaction parameter can be extracted from the statistical properties of the light emerging from the slab of nonlinear material, in particular in the real-space intensity-correlation signal (Sec.~\ref{SubSubSec:CorrelationsInPositionSpace}) and in the far-field angular distribution and two-body correlations of the transmitted light (Sec.~\ref{SubSubSec:CorrelationsInMomentumSpace}). An intuitive interpretation of our predictions in terms of a dynamical Casimir emission of Bogoliubov waves on top of a temporally modulated quantum gas of weakly interacting photons is provided and explained in detail, and the analogies and differences with standard four-wave-mixing experiments are discussed.

While working with spatially finite, e.g., Gaussian, beams and extending the whole theory to the general case where $U(\mathbf{x})\propto\delta\chi(\mathbf{x})\neq0$ and $g\propto\chi^{(3)}$ depends on $\mathbf{x}$ [cf.~Eq.~\eqref{Susceptibility}] is computationally demanding but conceptually straightforward, for the sake of simplicity we will restrict our attention to homogeneous geometries: In an actual experimental implementation, this assumption requires working with continuous-wave fields with a wide top-hat spatial profile propagating in a homogeneous medium, so that a uniform bulk region can be identified in the fluid of light, as done in the recent experiment \cite{Vocke2015}.

\subsection{Physical situation}
\label{SubSec:PhysicalSituation}

\begin{figure}
\includegraphics[width=\linewidth]{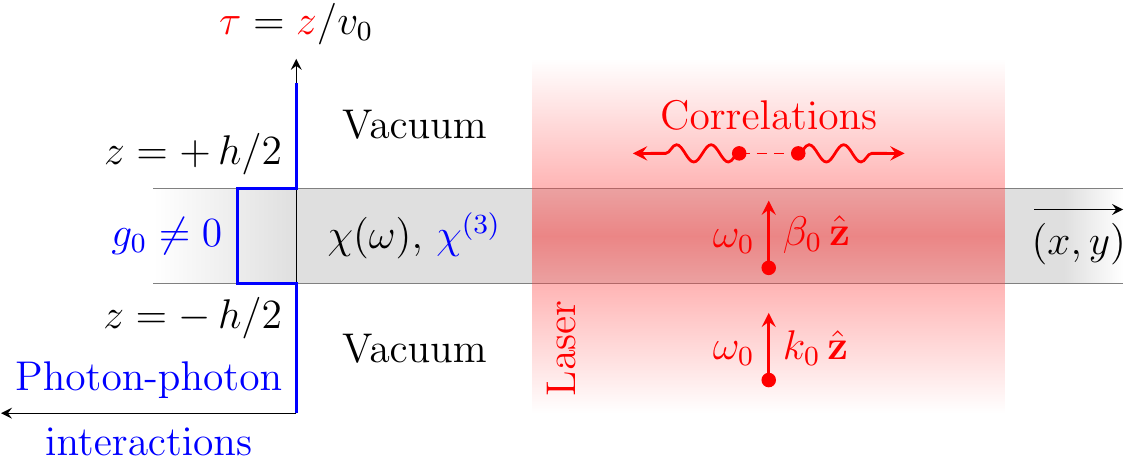}
\caption{(Color online) Sketch of the investigated configuration. A plate-shaped nonlinear medium of homogeneous linear electric susceptibility $\chi(\omega)$, constant Kerr coefficient $\chi^{(3)}$, and finite thickness $h$ in the $z$ direction is immersed in vacuum and illuminated by a wide monochromatic-plane-wave laser beam parallel to the $z$ axis. In such a configuration, the photon-photon interaction constant undergoes a sudden gate-shaped modulation along the optical axis. The resulting quantum correlations appearing in the transmitted light reveal the spontaneous emission of correlated counterpropagating excitations in the photon fluid.}
\label{ExperimentalSetup}
\end{figure}

As sketched in Fig.~\ref{ExperimentalSetup}, the slab of nonlinear material consists in a homogeneous dielectric layer of linear electric susceptibility $\chi(\omega)$ and uniform Kerr coefficient $\chi^{(3)}$. It is assumed to be infinite in the $x$ and $y$ directions and to have a uniform thickness $h$ in the $z$ one. The $z=-h/2$ ($z=h/2$) surface defines the front (back) interface between the dielectric and vacuum. The monochromatic laser beam of angular frequency $\omega_{0}$ which illuminates perpendicularly the medium is supposed to have a wide uniform profile in the transverse $(x,y)$ plane so that it can be legitimately seen as an infinite plane wave. The propagation occurs along the increasing-$z$ axis and the propagation constant in vacuum is denoted as $k_{0}=\omega_{0}/c$. At the angular frequency $\omega_{0}$, the material is supposed to have an anomalous group-velocity dispersion ($D_{0}<0$) and a negative Kerr coefficient [$\chi^{(3)}<0$, i.e., $g_{0}>0$; see Eq.~\eqref{UniformSituation2}], so that the fluid of light is dynamically stable inside the nonlinear medium (see Sec.~\ref{SubSec:SpatiallyHomogeneousSystem}).

Back-propagating light waves originating from reflection on the interfaces separating the Kerr material from vacuum would spoil the reformulation of the laser propagation in the positive-$z$ direction in terms of an effective time evolution. In order to avoid dealing with their existence, we assume that the $z=\mp\,h/2$ surfaces of the nonlinear-dielectric layer are treated with a perfect antireflection coating. Since its characteristic thickness in the $z$ direction (of the order of a few optical wavelengths) is typically much shorter than all the other lengths of the problem, its effect on light transmission can be summarized as a simple boundary condition guaranteeing the conservation of the energy-flux density of the electromagnetic wave---the so-called Poynting vector---at the $z=\mp\,h/2$ interfaces of the dielectic slab. Mathematically speaking, this may be explicitly expressed as
\begin{align}
\notag
&\left.\hat{\mathcal{E}}(\mathbf{x}_{\perp},z=\mp\,h/2\mp\epsilon,t)\,e^{\mp ik_{0}h/2}\right. \\
\label{ContinuityPoyntingVector}
&\left.=(1+\chi_{0})^{1/4}\,\hat{\mathcal{E}}(\mathbf{x}_{\perp},z=\mp\,h/2\pm\epsilon,t)\,e^{\mp i\beta_{0}h/2},\right.
\end{align}
where $\epsilon$ is an arbitrarily-small positive parameter. Equation \eqref{ContinuityPoyntingVector} allows for matching the quantum fluctuations of the optical field in vacuum ($|z|>h/2$) to the ones propagating in the bulk nonlinear medium ($|z|<h/2$); see Sec.~\ref{SubSubSec:Matching}.

\subsection{Time evolution of the quantum fluctuations}
\label{SubSec:TimeEvolutionOfTheQuantumFluctuations}

\subsubsection{Electric field for \texorpdfstring{$z<-h/2$}{Lg} and \texorpdfstring{$z>h/2$}{Lg}}
\label{SubSubSec:Outside}

Outside the Kerr layer (when $|z|>h/2$), that is, in vacuum, the (quantized) electric field $\hat{E}(\mathbf{x},t)$ of the laser wave admits the usual Fourier expansion (see, e.g., Ref.~\cite{CohenTannoudji1989})
\begin{equation}
\label{ElectricFieldVacuum}
\hat{E}(\mathbf{x},t)=i\int\frac{d\mathbf{k}}{(2\pi)^{3}}\,\sqrt{\frac{\hbar\,\omega(\mathbf{k})}{2\,\varepsilon_{0}}}\,e^{i[\mathbf{k}\cdot\mathbf{x}-\omega(\mathbf{k})t]}\,\hat{\alpha}(\mathbf{k})+\mathrm{H.c.},
\end{equation}
where ``H.c.'' stands for ``Hermitian conjugate.'' In this equation, $\omega(\mathbf{k})=c\,|\mathbf{k}|$ denotes the photon dispersion law in vacuum, (we recall that) $\varepsilon_{0}$ is the vacuum permittivity, and the $\hat{\alpha}(\mathbf{k})$'s [$\hat{\alpha}^{\dag}(\mathbf{k})$'s] are the photon destruction (creation) operators in the state of wavevector $\mathbf{k}=(\mathbf{k}_{\perp},k_{z})=(k_{x},k_{y},k_{z})$, subject to the boson commutation relations
\begin{subequations}
\label{PhotonCommutationRelations}
\begin{align}
\label{PhotonCommutationRelations1}
&\left.[\hat{\alpha}(\mathbf{k}),\hat{\alpha}^{\dag}(\mathbf{k}')]=(2\pi)^{3}\,\delta(\mathbf{k}-\mathbf{k}')\right. \\
\label{PhotonCommutationRelations2}
\text{and}\quad&\left.[\hat{\alpha}(\mathbf{k}),\hat{\alpha}(\mathbf{k}')]=0.\right.
\end{align}
\end{subequations}

In vacuum, the second-quantized version of Eq.~\eqref{ElectricField} reads $\hat{E}(\mathbf{x},t)=\frac{1}{2}\,\hat{\mathcal{E}}(\mathbf{x},t)\,e^{i(k_{0}z-\omega_{0}t)}+\mathrm{H.c.}$, where $k_{0}=\omega_{0}/c$ is the propagation constant of the light wave along the positive-$z$ direction in free space. Thus, by identification, one deduces from Eq.~\eqref{ElectricFieldVacuum} that the envelope $\hat{\mathcal{E}}(\mathbf{x},t)$ of the electric field in vacuum is given by
\begin{align}
\notag
\hat{\mathcal{E}}(\mathbf{x},t)=&\left.i\int\frac{d\mathbf{k}_{\perp}\,d\delta k_{z}}{(2\pi)^{3}}\,\sqrt{\frac{2\,\hbar\,[\omega_{0}+\delta\omega(\mathbf{k}_{\perp},\delta k_{z})]}{\varepsilon_{0}}}\right. \\
\label{EnvelopeVacuum}
&\left.\times\,e^{i[\mathbf{k}_{\perp}\cdot\mathbf{x}_{\perp}+\delta k_{z}z-\delta\omega(\mathbf{k}_{\perp},\delta k_{z})t]}\,\hat{\alpha}(\mathbf{k}_{\perp},\delta k_{z}),\right.
\end{align}
equation in which we made the variable change $\delta k_{z}=k_{z}-k_{0}$ and in which we defined $\delta\omega(\mathbf{k}_{\perp},\delta k_{z})=\omega(\mathbf{k})-\omega_{0}$ and $\hat{\alpha}(\mathbf{k}_{\perp},\delta k_{z})=\hat{\alpha}(\mathbf{k}_{\perp},k_{0}+\delta k_{z})$. The $\hat{\alpha}(\mathbf{k}_{\perp},\delta k_{z})$'s obey the commutation rules \eqref{PhotonCommutationRelations}, assuming the exchanges $k_{z}\longleftrightarrow\delta k_{z}$ and $k_{z}'\longleftrightarrow\delta k_{z}'$. Furthermore, $\mathbf{k}_{\perp}$, $\delta k_{z}$, and $\delta\omega$ in Eq.~\eqref{EnvelopeVacuum} are related through
\begin{equation}
\label{DispersionVacuum}
\delta k_{z}(\mathbf{k}_{\perp},\delta\omega)=-\frac{\mathbf{k}_{\perp}^{2}}{2\,k_{0}}+\frac{\delta\omega}{c},
\end{equation}
which is obtained by expanding at leading order the vacuum dispersion law $\omega(\mathbf{k})=c\,|\mathbf{k}|$ around $(\mathbf{k}_{\perp},k_{z},\omega)=(0,k_{0},\omega_{0})$, that is, within the framework of the paraxial ($|\mathbf{k}_{\perp}|/k_{0}\ll1$) and slowly-varying-envelope ($|\delta k_{z}|/k_{0}\ll1$ and $|\delta\omega|/\omega_{0}\ll1$) approximations considered in this paper. Taking advantage of the dispersion law \eqref{DispersionVacuum}, one can transform the integration over $(\mathbf{k}_{\perp},\delta k_{z})$ in Eq.~\eqref{EnvelopeVacuum} into an integration over $(\mathbf{k}_{\perp},\delta\omega)$, yielding, at leading order in the paraxial and slowly-varying-envelope approximations,
\begin{align}
\notag
\hat{\mathcal{E}}(\mathbf{x},t)=&\left.i\,\sqrt{\frac{2\,\hbar\,\omega_{0}}{c^{2}\,\varepsilon_{0}}}\int\frac{d\mathbf{k}_{\perp}\,d\delta\omega}{(2\pi)^{3}}\right. \\
\label{FluctuationVacuum}
&\left.\times\,e^{i[\mathbf{k}_{\perp}\cdot\mathbf{x}_{\perp}+\delta k_{z}(\mathbf{k}_{\perp},\delta\omega)z-\delta\omega t]}\,\hat{\alpha}[\mathbf{k}_{\perp},\delta k_{z}(\mathbf{k}_{\perp},\delta\omega)].\right.
\end{align}
In this way, the expression of the electric-field envelope outside the nonlinear medium is easily inserted in the $z\longleftrightarrow t$ mapping language discussed in the previous sections.

Defining in the integral \eqref{FluctuationVacuum}
\begin{subequations}
\label{TimeEvolutionReformulation}
\begin{align}
\label{TimeEvolutionReformulation1}
&\left.(\mathbf{r},\tau)=(\mathbf{x}_{\perp},\zeta,\tau)=(\mathbf{x}_{\perp},v_{0}\,t,z/v_{0}),\right. \\
\label{TimeEvolutionReformulation2}
&\left.\mathbf{q}=(\mathbf{q}_{\perp},q_{\zeta})=(\mathbf{k}_{\perp},-\delta\omega/v_{0}),\right. \\
\label{TimeEvolutionReformulation3}
\text{and}\quad&\left.\hat{a}(\mathbf{q})=\sqrt{\frac{\hbar\,v_{0}}{\mathcal{N}_{\mathrm{vac}}}}\,\hat{\alpha}[\mathbf{k}_{\perp},\delta k_{z}(\mathbf{k}_{\perp},\delta\omega)],\right.
\end{align}
\end{subequations}
where $\mathcal{N}_{\mathrm{vac}}=k_{0}^{\phantom{2}}/(\mu_{0}^{\phantom{2}}\,\omega_{0}^{2})=1/(c\,\mu_{0}\,\omega_{0})$ denotes the vacuum value of the normalization constant \eqref{NormalizationConstant}, one easily shows that Eq.~\eqref{FluctuationVacuum} can be put into the form
\begin{subequations}
\label{FluctuationVacuumBis}
\begin{align}
\label{FluctuationVacuumBis1}
\hat{\mathcal{E}}(\mathbf{r},\tau)&=i\,\sqrt{\frac{2\,\omega_{0}\,v_{0}\,\mathcal{N}_{\mathrm{vac}}}{c^{2}\,\varepsilon_{0}}}\int\frac{d\mathbf{q}}{(2\pi)^{3}}\,e^{i\mathbf{q}\cdot\mathbf{r}}\,\hat{a}(\mathbf{q},\tau) \\
\label{FluctuationVacuumBis2}
&=i\,\sqrt{\frac{2\,v_{0}}{c}}\int\frac{d\mathbf{q}}{(2\pi)^{3}}\,e^{i\mathbf{q}\cdot\mathbf{r}}\,\hat{a}(\mathbf{q},\tau),
\end{align}
\end{subequations}
where, introducing the input/output quantum mode operators $\hat{a}_{\mathrm{in}/\mathrm{out}}(\mathbf{q})=\exp\{(+/-)\,i\,[\mathbf{q}_{\perp}^{2}/(2\,k_{0})+v_{0}\,q_{\zeta}/c]\,h/2\}\,\hat{a}(\mathbf{q})$,
\begin{align}
\notag
\hat{a}\big(\mathbf{q},\tau\lessgtr\mp\tfrac{h}{2\,v_{0}}\big)=&\left.\exp\bigg[{-}i\,v_{0}\,\bigg(\frac{\mathbf{q}_{\perp}^{2}}{2\,k_{0}}+\frac{v_{0}}{c}\,q_{\zeta}\bigg)\,\big(\tau\pm\tfrac{h}{2\,v_{0}}\big)\bigg]\right. \\
\label{LadderOperatorsInOut}
&\left.\times\,\hat{a}_{\left(\begin{smallmatrix}\mathrm{in}\\\mathrm{out}\end{smallmatrix}\right)}(\mathbf{q}).\right.
\end{align}
By construction, the ladder operators $\hat{a}(\mathbf{q},\tau)$ satisfy the equal-$\tau$ commutation relations [use Eqs.~\eqref{PhotonCommutationRelations} and \eqref{TimeEvolutionReformulation3}]
\begin{subequations}
\label{PhotonCommutationRelationsBis}
\begin{align}
\label{PhotonCommutationRelationsBis1}
&\left.[\hat{a}(\mathbf{q},\tau),\hat{a}^{\dag}(\mathbf{q}',\tau)]=(2\pi)^{3}\,\frac{\hbar\,c}{\mathcal{N}_{\mathrm{vac}}}\,\delta(\mathbf{q}-\mathbf{q}')\right. \\
\label{PhotonCommutationRelationsBis2}
\text{and}\quad&\left.[\hat{a}(\mathbf{q},\tau),\hat{a}(\mathbf{q}',\tau)]=0.\right.
\end{align}
\end{subequations}

In order to keep some homogeneity in the notations used outside and inside the nonlinear material, the coordinate referencing a point along the $t$ axis is conveniently defined in vacuum in the same way as in the Kerr layer: $\zeta=v_{0}\,t$ [Eq.~\eqref{TimeEvolutionReformulation1}], where $v_{0}$ denotes the group velocity in the Kerr medium (i.e., in the $|z|<h/2$ region) at the laser's angular frequency $\omega_{0}$. Finally, note that the $\hat{a}_{\mathrm{in}}(\mathbf{q})$'s [$\hat{a}_{\mathrm{out}}(\mathbf{q})$'s] are related to the usual input (output) photon operators $\hat{\alpha}_{\mathrm{in}}(\mathbf{k})$'s [$\hat{\alpha}_{\mathrm{out}}(\mathbf{k})$'s] in the $z<-h/2$ ($z>h/2$) vacuum through Eq.~\eqref{TimeEvolutionReformulation3}.

In the specific configuration investigated here, with a coherent monochromatic-plane-wave incident light beam, the optical field in vacuum can be expanded as
\begin{equation}
\label{LaserContribution}
\hat{\mathcal{E}}\big(\mathbf{r},\tau\lessgtr\mp\tfrac{h}{2\,v_{0}}\big)=\mathcal{E}_{0}\,e^{i\varphi_{\lessgtr}}+\delta\hat{\mathcal{E}}(\mathbf{r},\tau),
\end{equation}
i.e., as a classical background field $\mathcal{E}_{0}\,e^{i\varphi_{\lessgtr}}$ (with $\mathcal{E}_{0},\varphi_{\lessgtr}=\mathrm{const}\in\mathbb{R}$) plus a small quantum-fluctuation term $\delta\hat{\mathcal{E}}(\mathbf{r},\tau)$. The latter may be decomposed according to the same plane-wave expansion as the one, Eq.~\eqref{FluctuationVacuumBis}, of the total electric-field envelope: The assumption of having a purely coherent incident beam translates into the condition that all the incident modes other than the coherent-pump one are in their vacuum state $|0_{\mathrm{in}}\rangle$, by definition such that $\hat{\alpha}_{\mathrm{in}}(\mathbf{k}\neq\mathbf{k}_{\mathrm{pump}})\,|0_{\mathrm{in}}\rangle=0$, which correspondingly reads in terms of the $\hat{a}_{\mathrm{in}}(\mathbf{q})$ operators as $\hat{a}_{\mathrm{in}}(\mathbf{q}\neq0)\,|0_{\mathrm{in}}\rangle=0$.

\subsubsection{Electric field for \texorpdfstring{$-h/2<z<h/2$}{Lg}}
\label{SubSubSec:Inside}

Inside the nonlinear medium (when $|z|<h/2$), the (quantized) electric-field envelope $\hat{\mathcal{E}}(\mathbf{r},\tau)$ evolves in time $\tau=z/v_{0}$ in the $\mathbf{r}=(\mathbf{x}_{\perp},\zeta=v_{0}\,t)$ space according to the Heisenberg equation of motion \eqref{HeisenbergEquationMotion} with the prescription \eqref{UniformSituation}.

Within the Bogoliubov weakly nonlinear regime, the background electric-field envelope $\mathcal{E}(\mathbf{r},\tau)$ is of constant amplitude $\mathcal{E}$ and of global phase $\varphi(\tau)=\varphi_{0}-v_{0}\,g_{0}\,\mathcal{E}^{2}\,(\tau-\tau_{0})$ (see Sec.~\ref{SubSec:SpatiallyHomogeneousSystem}), where $\tau_{0}=-h/(2\,v_{0})$ is the time which (naturally) initializes the evolution of the optical field in the medium. From the continuity of the Poynting vector at the entrance face of the nonlinear material, i.e., from Eq.~\eqref{ContinuityPoyntingVector} for $z=-h/2$ [$\tau=-h/(2\,v_{0})=\tau_{0}$], one finds
\begin{equation}
\label{ContinuityConditionsBackground1}
\mathcal{E}=\frac{\mathcal{E}_{0}}{(1+\chi_{0})^{1/4}}\quad\text{and}\quad\varphi_{0}=\varphi_{<}+(\beta_{0}-k_{0})\,\frac{h}{2}.
\end{equation}
On the other hand, from the continuity of the Poynting vector at the exit face of the medium, i.e., from Eq.~\eqref{ContinuityPoyntingVector} for $z=h/2$ [$\tau=h/(2\,v_{0})$], one gets that the phase $\varphi_{>}$ of the optical field in the $z>h/2$ vacuum is locked at, using Eqs.~\eqref{ContinuityConditionsBackground1},
\begin{subequations}
\label{ContinuityConditionsBackground2}
\begin{align}
\label{ContinuityConditionsBackground21}
\varphi_{>}&=\varphi_{0}-g_{0}\,\mathcal{E}^{2}\,h+(\beta_{0}-k_{0})\,\frac{h}{2} \\
\label{ContinuityConditionsBackground22}
&=\varphi_{<}+(\beta_{0}-k_{0}-g_{0}\,\mathcal{E}^{2})\,h.
\end{align}
\end{subequations}

Finally, the small quantum fluctuation $\delta\hat{\mathcal{E}}(\mathbf{r},\tau)$ which superimposes upon the mean-field solution $\mathcal{E}(\mathbf{r},\tau)=\mathcal{E}\,e^{i\varphi(\tau)}$ is described by Eqs.~\eqref{BogoliubovPlaneWaves}--\eqref{BogoliubovAmplitudes}, with $\mathcal{E}$ and $\varphi_{0}$ satisfying the constraints \eqref{ContinuityConditionsBackground1}.

\subsubsection{Matching at \texorpdfstring{$z=-h/2$}{Lg} and \texorpdfstring{$z=h/2$}{Lg}}
\label{SubSubSec:Matching}

The relation between the fluctuations after ($z>h/2$) and before ($z<-h/2$) the gate-shaped modulation of the optical nonlinearity along the radiation axis are obtained by matching Eqs.~\eqref{BogoliubovPlaneWaves} and (\ref{FluctuationVacuumBis}, \ref{LaserContribution}) at the $z=\mp\,h/2$ interfaces using the continuity equation \eqref{ContinuityPoyntingVector}.

At the entrance face of the dielectric plate, that is, at $z=-h/2$ [$\tau=-h/(2\,v_{0})=\tau_{0}$], one has
\begin{align}
\notag
i\,\sqrt{\frac{2\,v_{0}}{c}}\,\hat{a}_{\mathrm{in}}(\mathbf{q})\,e^{-ik_{0}h/2}=&\left.(1+\chi_{0})^{1/4}\,e^{i\varphi_{0}}\,\Big[u_{\mathbf{q}}\,\hat{b}(\mathbf{q},\tau_{0})\right.\\
\label{MatchingInFace}
&\left.+\,v_{\mathbf{q}}\,\hat{b}^{\dag}(-\mathbf{q},\tau_{0})\Big]\,e^{-i\beta_{0}h/2},\right.
\end{align}
in such a way that, taking advantage of $u_{\mathbf{q}}^{2}-v_{\mathbf{q}}^{2}=1$ and of the second of Eqs.~\eqref{ContinuityConditionsBackground1},
\begin{align}
\notag
\hat{b}(\mathbf{q},\tau_{0})=&\left.i\,\sqrt{\frac{2\,v_{0}/c}{\sqrt{1+\chi_{0}}}}\right. \\
\label{MatchingInFaceBis}
&\left.\times\,\Big[e^{-i\varphi_{<}}\,u_{\mathbf{q}}\,\hat{a}_{\mathrm{in}}^{\phantom{\dag}}(\mathbf{q})+e^{i\varphi_{<}}\,v_{\mathbf{q}}\,\hat{a}_{\mathrm{in}}^{\dag}(-\mathbf{q})\Big].\right.
\end{align}
At the back face of the Kerr layer, that is, at $z=h/2$ [$\tau=h/(2\,v_{0})$], the continuity of the Poynting vector yields
\begin{align}
\notag
&\left.i\,\sqrt{\frac{2\,v_{0}}{c}}\,\hat{a}_{\mathrm{out}}(\mathbf{q})\,e^{ik_{0}h/2}\right. \\
\notag
&\left.=(1+\chi_{0})^{1/4}\,e^{i(\varphi_{0}-g_{0}\mathcal{E}^{2}h)}\,\Big[u_{\mathbf{q}}\,e^{-i\omega_{\mathbf{q}}h/v_{0}}\,\hat{b}(\mathbf{q},\tau_{0})\right. \\
\label{MatchingOutFace}
&\left.\phantom{=}+v_{\mathbf{q}}\,e^{i\omega_{-\mathbf{q}}h/v_{0}}\,\hat{b}^{\dag}(-\mathbf{q},\tau_{0})\Big]\,e^{i\beta_{0}h/2},\right.
\end{align}
from which and using $\omega_{\mathbf{q}}=v_{0}\,[K_{\mathrm{B}}(\mathbf{q})+q_{\zeta}]$, Eq.~\eqref{ContinuityConditionsBackground21}, and Eq.~\eqref{MatchingInFaceBis} one finally obtains
\begin{equation}
\label{InOutRelation}
\hat{a}_{\mathrm{out}}^{\phantom{\dag}}(\mathbf{q})=\tilde{u}_{\mathbf{q}}^{\phantom{\ast}}\,\hat{a}_{\mathrm{in}}^{\phantom{\dag}}(\mathbf{q})+\tilde{v}_{-\mathbf{q}}^{\ast}\,\hat{a}_{\mathrm{in}}^{\dag}(-\mathbf{q}),
\end{equation}
where we have defined
\begin{subequations}
\label{NewBogoliubovAmplitudes}
\begin{align}
\label{NewBogoliubovAmplitudes1}
&\left.\tilde{u}_{\mathbf{q}}=e^{-i(\varphi_{<}-\varphi_{>}+q_{\zeta}h)}\,\mathcal{U}_{\mathbf{q}}\right. \\
\label{NewBogoliubovAmplitudes2}
\text{and}\quad&\left.\tilde{v}_{\mathbf{q}}=-e^{-i(\varphi_{<}+\varphi_{>}+q_{\zeta}h)}\,\mathcal{V}_{\mathbf{q}},\right.
\end{align}
\end{subequations}
with
\begin{subequations}
\label{NewBogoliubovAmplitudesBis}
\begin{align}
\label{NewBogoliubovAmplitudesBis1}
&\left.\mathcal{U}_{\mathbf{q}}=u_{\mathbf{q}}^{2}\,e^{-iK_{\mathrm{B}}(\mathbf{q})h}-v_{\mathbf{q}}^{2}\,e^{iK_{\mathrm{B}}(\mathbf{q})h}\right. \\
\label{NewBogoliubovAmplitudesBis2}
\text{and}\quad&\left.\mathcal{V}_{\mathbf{q}}=u_{\mathbf{q}}\,v_{\mathbf{q}}\,\Big[e^{-iK_{\mathrm{B}}(\mathbf{q})h}-e^{iK_{\mathrm{B}}(\mathbf{q})h}\Big].\right.
\end{align}
\end{subequations}

\subsection{Statistical properties of the transmitted light}
\label{SubSec:StatisticalPropertiesOfTheTransmittedLight}

From the input-output relation \eqref{InOutRelation}, it is immediate to extract predictions for all the coherence properties of the transmitted light field \eqref{LaserContribution} past the slab of nonlinear medium. As already mentioned in the end of Sec.~\ref{SubSubSec:Outside}, our assumption of a perfectly coherent pump incident on the dielectric corresponds to having all the fluctuation modes of the photon field in the vacuum state $|0_{\mathrm{in}}\rangle$ [$\hat{a}_{\mathrm{in}}(\mathbf{q}\neq0)\,|0_{\mathrm{in}}\rangle=0$]. As a consequence, denoting $\langle\cdot\rangle=\langle0_{\mathrm{in}}|\cdot|0_{\mathrm{in}}\rangle$ the quantum average in the state $|0_{\mathrm{in}}\rangle$, one initially has, just before entering the nonlinear material,
\begin{equation}
\label{AverageOfInterest1}
\langle\hat{a}_{\mathrm{in}}(\mathbf{q})\,\hat{a}_{\mathrm{in}}(\mathbf{q}')\rangle=\langle\hat{a}_{\mathrm{in}}^{\dag}(\mathbf{q})\,\hat{a}_{\mathrm{in}}^{\phantom{\dag}}(\mathbf{q}')\rangle=0
\end{equation}
and, thanks to the commutation relation \eqref{PhotonCommutationRelationsBis1},
\begin{equation}
\label{AverageOfInterest2}
\langle\hat{a}_{\mathrm{in}}^{\phantom{\dag}}(\mathbf{q})\,\hat{a}_{\mathrm{in}}^{\dag}(\mathbf{q}')\rangle=(2\pi)^{3}\,\frac{\hbar\,c}{\mathcal{N}_{\mathrm{vac}}}\,\delta(\mathbf{q}-\mathbf{q}').
\end{equation}

\subsubsection{Correlations in position space}
\label{SubSubSec:CorrelationsInPositionSpace}

While the average light intensity remains equal to the incident one in vacuum, the effect of the quenches of the photon-photon interaction constant along the optical axis is visible in the intensity-intensity correlation function
\begin{equation}
\label{IntensityCorrelationsDefinition}
g^{(2)}(\mathbf{r},\mathbf{r}';\tau)=\frac{\langle\,{:}\,\hat{\mathcal{I}}(\mathbf{r},\tau)\,\hat{\mathcal{I}}(\mathbf{r}',\tau)\,{:}\,\rangle}{\langle\hat{\mathcal{I}}(\mathbf{r},\tau)\rangle\,\langle\hat{\mathcal{I}}(\mathbf{r}',\tau)\rangle}-1
\end{equation}
at a fixed point $z>h/2$ of the laser-beam axis, i.e., at a fixed time $\tau>h/(2\,v_{0})$ after the quench sequence. In Eq.~\eqref{IntensityCorrelationsDefinition}, $\hat{\mathcal{I}}(\mathbf{r},\tau)=\hat{\mathcal{E}}^{\dag}(\mathbf{r},\tau)\,\hat{\mathcal{E}}(\mathbf{r},\tau)$ is the (quantum) intensity operator at $(\mathbf{r},\tau)=(\mathbf{x}_{\perp},\zeta,\tau)=(x,y,v_{0}\,t,z/v_{0})$ and the colon symbol denotes the normal-ordering operation.

Since we are dealing with a strongly coherent light wave, the electric-field envelope $\hat{\mathcal{E}}(\mathbf{r},\tau)$ at a time $\tau>h/(2\,v_{0})$ after the second quench of the photon-photon interaction constant is given by the expansion \eqref{LaserContribution}, where the weak quantum modulation $\delta\hat{\mathcal{E}}(\mathbf{r},\tau)$ obeys the plane-wave decomposition \eqref{FluctuationVacuumBis}. At leading order in the small fluctuations $\delta\hat{\mathcal{E}}(\mathbf{r},\tau)$ and $\delta\hat{\mathcal{E}}^{\dag}(\mathbf{r},\tau)$, the intensity operator $\hat{\mathcal{I}}(\mathbf{r},\tau)$ may be expressed as
\begin{subequations}
\label{IntensityOperatorBogoliubov}
\begin{align}
\label{IntensityOperatorBogoliubov1}
&\left.\hat{\mathcal{I}}(\mathbf{r},\tau)=\mathcal{E}_{0}^{2}+\delta\hat{\mathcal{I}}(\mathbf{r},\tau),\right. \\
\label{IntensityOperatorBogoliubov2}
\text{where}\quad&\left.\delta\hat{\mathcal{I}}(\mathbf{r},\tau)=\mathcal{E}_{0}\,e^{-i\varphi_{>}}\,\delta\hat{\mathcal{E}}(\mathbf{r},\tau)+\mathrm{H.c.}\right.
\end{align}
\end{subequations}
is a small quantum correction to the background light intensity $\mathcal{E}_{0}^{2}$ outside the nonlinear medium. Taking advantage of Eqs.~\eqref{FluctuationVacuumBis} and \eqref{PhotonCommutationRelationsBis1}, the intensity correlator \eqref{IntensityCorrelationsDefinition} becomes, at leading order in the expansion \eqref{IntensityOperatorBogoliubov1},
\begin{equation}
\label{IntensityCorrelationsDefinitionBis}
g^{(2)}(\mathbf{r},\mathbf{r}';\tau)=\frac{\langle\delta\hat{\mathcal{I}}(\mathbf{r},\tau)\,\delta\hat{\mathcal{I}}(\mathbf{r}',\tau)\rangle}{\mathcal{E}_{0}^{4}}-\frac{2\,v_{0}}{c}\,\frac{\hbar\,\omega_{0}}{\varepsilon_{0}^{\phantom{2}}\,\mathcal{E}_{0}^{2}}\,\delta(\mathbf{r}-\mathbf{r}').
\end{equation}
Physically, the second term in the right-hand side of Eq.~\eqref{IntensityCorrelationsDefinitionBis} is the shot-noise term originating from the discreteness, i.e., the quantum nature, of the photon in vacuum; mathematically, it is the consequence of the normal order considered in Eq.~\eqref{IntensityCorrelationsDefinition}.

Straightforward manipulations using Eqs.~\eqref{FluctuationVacuumBis}, \eqref{LadderOperatorsInOut}, \eqref{InOutRelation}, \eqref{NewBogoliubovAmplitudes}, \eqref{AverageOfInterest1}, \eqref{AverageOfInterest2}, and \eqref{IntensityOperatorBogoliubov2} lead to
\begin{align}
\notag
g^{(2)}(\mathbf{r},\mathbf{r}';\tau)=&\left.\frac{2\,v_{0}}{c}\,\frac{\hbar\,\omega_{0}}{\varepsilon_{0}^{\phantom{2}}\,\mathcal{E}_{0}^{2}}\int\frac{d\mathbf{q}}{(2\pi)^{3}}\,e^{i\mathbf{q}\cdot(\mathbf{r}-\mathbf{r}')}\right. \\
\notag
&\left.\times\,\bigg\{\bigg|\mathcal{U}_{\mathbf{q}}\exp\bigg[{-}i\,\frac{\mathbf{q}_{\perp}^{2}}{2\,k_{0}}\,\big(v_{0}\,\tau-\tfrac{h}{2}\big)\bigg]\right. \\
\label{IntensityCorrelations}
&\left.+\,\mathcal{V}_{\mathbf{q}}\exp\bigg[i\,\frac{\mathbf{q}_{\perp}^{2}}{2\,k_{0}}\,\big(v_{0}\,\tau-\tfrac{h}{2}\big)\bigg]\bigg|^{2}-1\bigg\}.\right.
\end{align}
The physics of these intensity correlations is most transparent at the exit face of the Kerr layer, i.e., at $\tau=h/(2\,v_{0})$. In this case, the correlation function \eqref{IntensityCorrelations} admits a very simple integral formulation; manipulating the $\mathcal{U}_{\mathbf{q}}$'s and the $\mathcal{V}_{\mathbf{q}}$'s given by Eqs.~\eqref{NewBogoliubovAmplitudesBis} and denoting $g^{(2)}(\mathbf{r}-\mathbf{r}')=g^{(2)}[\mathbf{r},\mathbf{r}';\tau=h/(2\,v_{0})]$, one indeed finds
\begin{subequations}
\label{IntensityCorrelationsExitFace}
\begin{align}
\notag
g^{(2)}(\mathbf{r}-\mathbf{r}')&\left.=\frac{2\,v_{0}}{c}\,\frac{\hbar\,\omega_{0}}{\varepsilon_{0}^{\phantom{2}}\,\mathcal{E}_{0}^{2}}\int\frac{d\mathbf{q}}{(2\pi)^{3}}\,e^{i\mathbf{q}\cdot(\mathbf{r}-\mathbf{r}')}\right. \\
\label{IntensityCorrelationsExitFace1}
&\left.\phantom{=}\times\bigg[\frac{K_{\mathrm{kin}}^{2}(\mathbf{q})}{K_{\mathrm{B}}^{2}(\mathbf{q})}-1\bigg]\sin^{2}[K_{\mathrm{B}}(\mathbf{q})\,h]\right. \\
\notag
&\left.=-\frac{2}{\pi^{2}}\,\frac{2\,v_{0}}{c}\,\frac{\hbar\,\omega_{0}}{\varepsilon_{0}^{\phantom{2}}\,\mathcal{E}_{0}^{2}\,\xi_{\perp}^{2}\,\xi_{\zeta}^{\phantom{2}}}\int_{0}^{\infty}d\kappa\;\mathrm{sinc}(\kappa\,\rho)\right. \\
\label{IntensityCorrelationsExitFace2}
&\left.\phantom{=}\times\frac{\kappa^{2}}{\kappa^{2}+4}\,\sin^{2}\bigg(\frac{\ell}{2}\,\kappa\,\sqrt{\kappa^{2}+4}\bigg),\right.
\end{align}
\end{subequations}
where $\xi_{\perp}=1/(\mathcal{M}_{x,x}\,g_{0}\,\mathcal{E}^{2})^{1/2}=1/(\mathcal{M}_{y,y}\,g_{0}\,\mathcal{E}^{2})^{1/2}$ and $\xi_{\zeta}=1/(\mathcal{M}_{\zeta,\zeta}\,g_{0}\,\mathcal{E}^{2})^{1/2}$ are the healing lengths in the $\mathbf{x}_{\perp}=(x,y)$ plane and in the $\zeta$ direction in the dielectric, with $\mathcal{E}^{2}=\mathcal{E}_{0}^{2}/(1+\chi_{0}^{\phantom{2}})^{1/2}$ the internal light intensity, $\ell=g_{0}\,\mathcal{E}^{2}\,h$ is the thickness of the nonlinear medium in units of $1/(g_{0}\,\mathcal{E}^{2})$, $\mathrm{sinc}(X)=\sin(X)/X$ is the so-called sine cardinal function, and
\begin{subequations}
\label{RelativeDistance}
\begin{align}
\label{RelativeDistance1}
\rho&=\sqrt{\frac{|\mathbf{x}_{\perp}^{\phantom{\prime}}-\mathbf{x}_{\perp}'|^{2}}{\xi_{\perp}^{2}}+\frac{|\zeta-\zeta'|^{2}}{\xi_{\zeta}^{2}}} \\
\label{RelativeDistance2}
&=\sqrt{\frac{|\mathbf{x}_{\perp}^{\phantom{\prime}}-\mathbf{x}_{\perp}'|^{2}}{\xi_{\perp}^{2}}+\frac{v_{0}^{2}\,|t-t'|^{2}}{\xi_{\zeta}^{2}}}
\end{align}
\end{subequations}
is the dimensionless relative spatiotemporal distance, $|\mathbf{x}_{\perp}^{\phantom{\prime}}-\mathbf{x}_{\perp}'|$ and $|\zeta-\zeta'|=v_{0}\,|t-t'|$ being naturally measured in units of $\xi_{\perp}$ and $\xi_{\zeta}$, respectively.

\begin{figure}
\includegraphics[width=\linewidth]{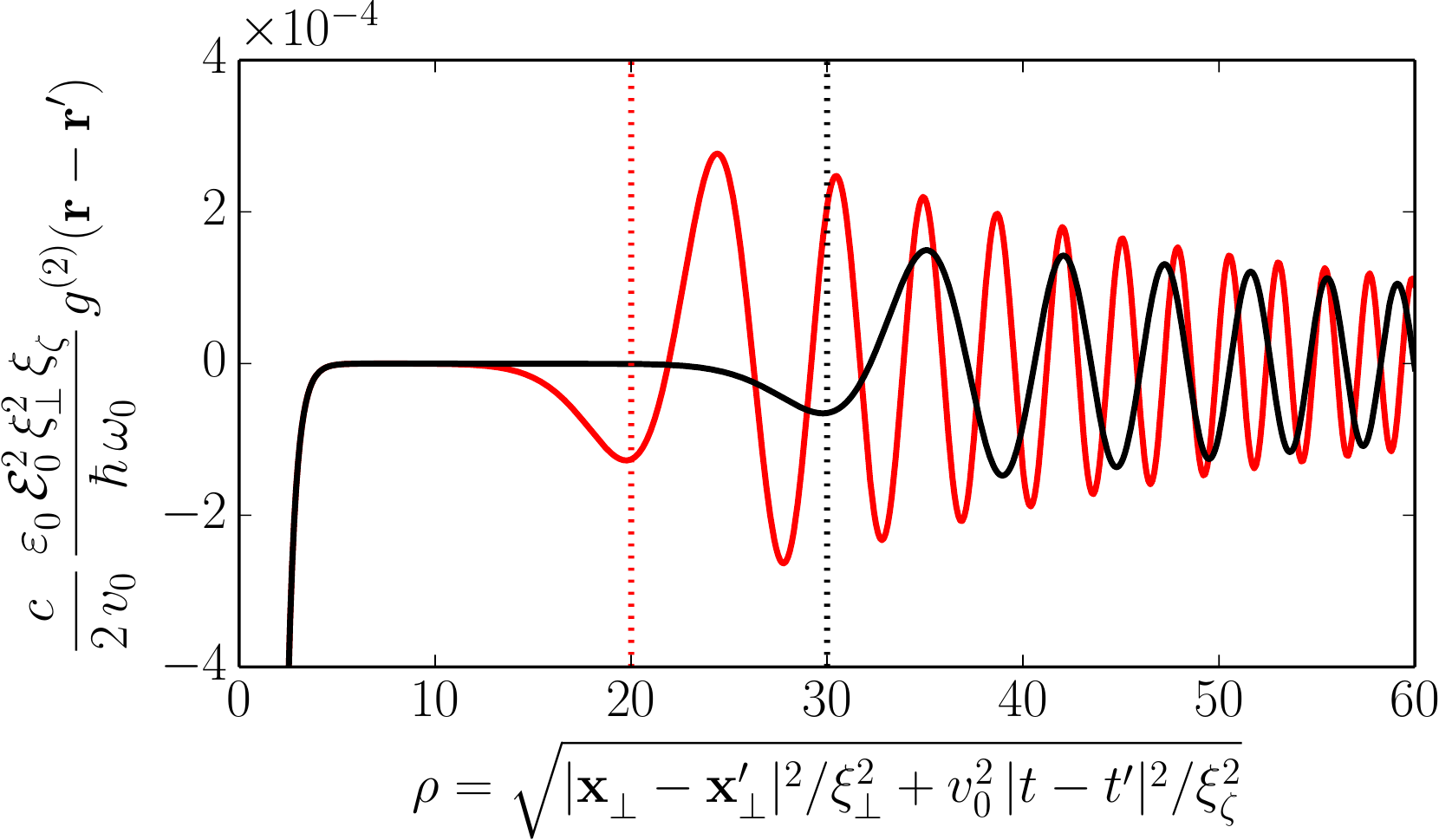}
\caption{(Color online) Normalized intensity-intensity correlator \eqref{IntensityCorrelationsExitFace} when $\ell=g_{0}\,\mathcal{E}^{2}\,h=10$ (red curve) and $\ell=15$ (black curve). The oscillation pattern appearing after $\rho_{\mathrm{c}}=2\,\ell=20$ (indicated by the red dotted line) or $\rho_{\mathrm{c}}=30$ (black dotted line) is the signature of the occurrence of a dynamical Casimir emission of elementary excitations in the quantum fluid of light.}
\label{Correlations}
\end{figure}

The graph of the two-body correlator $g^{(2)}$ [Eqs.~\eqref{IntensityCorrelationsExitFace}] as a function of the dimensionless relative distance $\rho$ [Eqs.~\eqref{RelativeDistance}] is represented in Fig.~\ref{Correlations}. As one can note from this plot, $g^{(2)}$ is characterized by a pronounced dip localized around $\rho=0$, that is, at equal $\mathbf{x}_{\perp}$ in the transverse plane and equal time $t$: $(\mathbf{x}_{\perp}^{\phantom{\prime}},t)=(\mathbf{x}_{\perp}',t')$. This antibunching is a direct consequence of the repulsive character ($g_{0}>0$) of the photon-photon interactions in the nonlinear medium and has been widely studied in the context of many-body physics (see, e.g., Refs.~\cite{Naraschewski1999, Carusotto2008, Deuar2009, Jacqmin2011, Larre2012}). Note that the nontrivial spatial structure of $g^{(2)}$ shown in Fig.~\ref{Correlations} makes the latter conceptually different from the single-mode antibunching behavior that is obtained, e.g., by resonantly scattering light off a two-level emitter \cite{CohenTannoudji1989} or by photon blockade in a strongly nonlinear cavity \cite{Carusotto2013}.

The most important feature of $g^{(2)}$ is the fringe pattern that is visible at larger $\rho$'s. The latter is concentrated in the $\rho\gtrsim2\,\ell$ region and can be interpreted from simple physical arguments \cite{Calabrese2006, Calabrese2007}. When the sudden change of the photon-photon interaction strength from zero to $g_{0}\neq0$ occurs at $z=-h/2$ [$\tau=-h/(2\,v_{0})=\tau_{0}$], pairs of correlated Bogoliubov waves of opposite wavevectors $\pm\,\mathbf{q}=\pm\,(\mathbf{q}_{\perp},q_{\zeta})=\pm\,(q_{x},q_{y},q_{\zeta})$ are spontaneously emitted in the $x$, $y$, and $\zeta$ directions. Inside the plate [that is, for $\tau_{0}<\tau<h/(2\,v_{0})$], a Bogoliubov excitation of wavevector $\mathbf{q}$ propagates at the group velocity ($\omega_{\mathbf{q}}=v_{0}\,[K_{\mathrm{B}}(\mathbf{q})+q_{\zeta}]$; see Sec.~\ref{SubSec:SpatiallyHomogeneousSystem})
\begin{subequations}
\label{GroupVelocity}
\begin{align}
\label{GroupVelocity1}
&\left.\mathbf{v}(\mathbf{q})=\frac{\partial\omega_{\mathbf{q}}}{\partial\mathbf{q}}=\mathbf{v}_{\perp}(\mathbf{q})+\mathbf{v}_{\zeta}(\mathbf{q}),\right. \\
\label{GroupVelocity2}
\text{where}\quad&\left.\mathbf{v}_{\perp}(\mathbf{q})=v_{0}\,\bigg[\frac{\partial K_{\mathrm{B}}}{\partial q_{x}}(\mathbf{q})\,\mathbf{u}_{x}+\frac{\partial K_{\mathrm{B}}}{\partial q_{y}}(\mathbf{q})\,\mathbf{u}_{y}\bigg]\right. \\
\label{GroupVelocity3}
\text{and}\quad&\left.\mathbf{v}_{\zeta}(\mathbf{q})=v_{0}\,\bigg[\frac{\partial K_{\mathrm{B}}}{\partial q_{\zeta}}(\mathbf{q})+1\bigg]\,\mathbf{u}_{\zeta},\right.
\end{align}
\end{subequations}
$\mathbf{u}_{x}$ ($\mathbf{u}_{y}$, $\mathbf{u}_{\zeta}$) being a unit vector in the $x$ ($y$, $\zeta$) direction. Thus, as $\mathbf{v}$ is a growing function of $\mathbf{q}$ [see the definition of $K_{\mathrm{B}}(\mathbf{q})$ in Eq.~\eqref{BogoliubovSpectrum}], the modes forming a pair of entangled excitations of opposite momenta must be separated at a time $\tau>\tau_{0}$ [and such that $\tau<h/(2\,v_{0})$] by a distance
\begin{subequations}
\label{TransverseSeparation}
\begin{align}
\label{TransverseSeparation1}
|\mathbf{x}_{\perp}^{\phantom{\prime}}-\mathbf{x}_{\perp}'|&\gtrsim\Big|[\mathbf{v}_{\perp}(\mathbf{q})-\mathbf{v}_{\perp}(-\mathbf{q})]_{\mathbf{q}\to0}\Big|\,(\tau-\tau_{0}) \\
\label{TransverseSeparation2}
&\simeq\frac{2\,s_{\perp}^{2}\,|\mathbf{q}_{\perp}^{\phantom{2}}|}{\sqrt{(s_{\perp}\,|\mathbf{q}_{\perp}|)^{2}+(s_{\zeta}\,q_{\zeta})^{2}}}\,v_{0}\,(\tau-\tau_{0})
\end{align}
\end{subequations}
in the $\mathbf{x}_{\perp}=(x,y)$ plane and by a distance
\begin{subequations}
\label{TemporalSeparation}
\begin{align}
\label{TemporalSeparation1}
|\zeta-\zeta'|&\gtrsim\Big|[\mathbf{v}_{\zeta}(\mathbf{q})-\mathbf{v}_{\zeta}(-\mathbf{q})]_{\mathbf{q}\to0}\Big|\,(\tau-\tau_{0}) \\
\label{TemporalSeparation2}
&\simeq\frac{2\,s_{\zeta}^{2}\,|q_{\zeta}^{\phantom{2}}|}{\sqrt{(s_{\perp}\,|\mathbf{q}_{\perp}|)^{2}+(s_{\zeta}\,q_{\zeta})^{2}}}\,v_{0}\,(\tau-\tau_{0})
\end{align}
\end{subequations}
along the $\zeta=v_{0}\,t$ direction, where $s_{\perp/\zeta}=g_{0}\,\mathcal{E}^{2}\,\xi_{\perp/\zeta}$ are the ``sound speeds'' (see Sec.~\ref{SubSec:SpatiallyHomogeneousSystem}) in the transverse and temporal directions inside the nonlinear medium. As a result, according to \eqref{TransverseSeparation} and \eqref{TemporalSeparation}, the (dimensionless) distance $\rho$ separating two correlated counterpropagating modes has to be
\begin{equation}
\label{DimensionlessSeparation}
\rho\gtrsim2\,g_{0}\,\mathcal{E}^{2}\,v_{0}\,(\tau-\tau_{0}),
\end{equation}
and so $\rho\gtrsim\rho_{\mathrm{c}}$, with
\begin{equation}
\label{CriticalRho}
\rho_{\mathrm{c}}=2\,g_{0}\,\mathcal{E}^{2}\,h=2\,\ell,
\end{equation}
at the back face of the nonlinear medium where $\tau=h/(2\,v_{0})$. At this point, the correlation between counterpropagating Bogoliubov excitations generated by the quench of the optical nonlinearity at $z=-h/2$ results in an oscillatory structure in the $g^{(2)}$ function that is concentrated in the $\rho\gtrsim\rho_{\mathrm{c}}$ region, which we easily verify on the examples of Fig.~\ref{Correlations} where $\ell=10$ (red curve) and $\ell=15$ (black curve). The amplitude of the fringe pattern diminishes as the separation distance $\rho$ grows: The correlations between Bogoliubov excitations of opposite momenta are naturally all the weaker as the excitations are separated in space and time.

This oscillatory behavior of the correlation function in position space can be interpreted as resulting from a dynamical Casimir emission \cite{Moore1970, Fulling1976, Davies1977, Kardar1999} of elementary excitations on top of a quantum fluid of light presenting a quench of the photon-photon interaction constant in time $\tau=z/v_{0}$. This has been theoretically studied in Ref.~\cite{Carusotto2010} in the case of a weakly interacting Bose-Einstein condensate presenting a quick time modulation of the atom-atom $s$-wave scattering length. From the experimental point of view, the closely related phenomenon of Sakharov oscillations \cite{Sakharov1965}, resulting from the interference of synchronously generated counterpropagating acoustic waves, has been recently observed in the density fluctuations of cesium superfluids after an abrupt quench in time of the atom interaction strength \textit{via} Feshbach resonances \cite{Hung2013}: This last observation is of particular interest given its connection to the temperature fluctuations in the cosmic microwave background radiation \cite{Schmiedmayer2013}.

\subsubsection{Correlations in momentum space}
\label{SubSubSec:CorrelationsInMomentumSpace}

The emission of correlated-phonon pairs in the fluid of light due to the two quantum quenches of the photon-photon interaction constant can also be detected in momentum space by taking a far-field and spectrally resolved picture of the light emerging from the back face of the nonlinear medium. In the optical language, this process of emission of photon pairs at wavevectors and frequencies different from the incident ones goes under the name of spontaneous four-wave mixing.

A simple measurement that could be done would be the one of the photon momentum distribution $\langle\hat{N}(\mathbf{q},\tau)\rangle$ at a fixed point $z>h/2$ of the radiation axis, that is, at a fixed time $\tau>h/(2\,v_{0})$, as a function of the wavevector $\mathbf{q}$. Reminding the definition of $\mathbf{q}$, i.e., $\mathbf{q}=(\mathbf{k}_{\perp},-\delta\omega/v_{0})$, this measurement involves (i) an angular resolution to isolate the light deflected with a transverse wavevector $\mathbf{k}_{\perp}$ and (ii) a spectral resolution to isolate the angular-frequency component of the transmitted light at $\omega=\omega_{0}\pm\delta\omega$. By construction, the population operator $\hat{N}(\mathbf{q},\tau)=\hat{a}^{\dag}(\mathbf{q},\tau)\,\hat{a}(\mathbf{q},\tau)$ in the state of wavevector $\mathbf{q}$ does not depend on time $\tau$ [see Eq.~\eqref{LadderOperatorsInOut}] and one simply has $\hat{N}(\mathbf{q},\tau)=\hat{a}_{\mathrm{out}}^{\dag}(\mathbf{q})\,\hat{a}_{\mathrm{out}}^{\phantom{\dag}}(\mathbf{q})$ for all $\tau>h/(2\,v_{0})$. Thus, using Eqs.~\eqref{InOutRelation}--\eqref{AverageOfInterest2}, one obtains
\begin{subequations}
\label{MomentumDistribution}
\begin{align}
\notag
&\left.\langle\hat{N}(\mathbf{q},\tau)\rangle\right. \\
\label{MomentumDistribution1}
&\left.=|\tilde{v}_{-\mathbf{q}}|^{2}\,(2\pi)^{3}\,\frac{\hbar\,c}{\mathcal{N}_{\mathrm{vac}}}\,\frac{A\,v_{0}\,\Delta t}{(2\pi)^{3}}\right. \\
\label{MomentumDistribution2}
&\left.=A\,v_{0}\,\Delta t\,\frac{\hbar\,\omega_{0}}{\varepsilon_{0}}\left[\frac{g_{0}\,\mathcal{E}^{2}}{K_{\mathrm{B}}(\mathbf{q})}\right]^{2}\sin^{2}[K_{\mathrm{B}}(\mathbf{q})\,h]\right. \\
\label{MomentumDistribution3}
&\left.=A\,v_{0}\,\Delta t\,\frac{\hbar\,\omega_{0}}{\varepsilon_{0}}\,\frac{4}{\kappa^{2}\,(\kappa^{2}+4)}\sin^{2}\bigg(\frac{\ell}{2}\,\kappa\,\sqrt{\kappa^{2}+4}\bigg),\right.
\end{align}
\end{subequations}
where $A$ and $\Delta t$ ($A,\Delta t\to\infty$) empirically denote the typical transverse cross section of the beam of light and the typical duration of the measurement \cite{NoteMomentumSpace}, respectively, $\ell=g_{0}\,\mathcal{E}^{2}\,h$ is, as before, the normalized slab thickness, and
\begin{equation}
\label{ExcitationWavenumber}
\kappa=\sqrt{\xi_{\perp}^{2}\,\mathbf{q}_{\perp}^{2}+\xi_{\zeta}^{2}\,q_{\zeta}^{2}}
\end{equation}
is the dimensionless total excitation wavenumber, $\mathbf{q}_{\perp}$ and $q_{\zeta}$ being naturally measured in units of $1/\xi_{\perp}$ and $1/\xi_{\zeta}$, respectively.

\begin{figure}
\includegraphics[width=\linewidth]{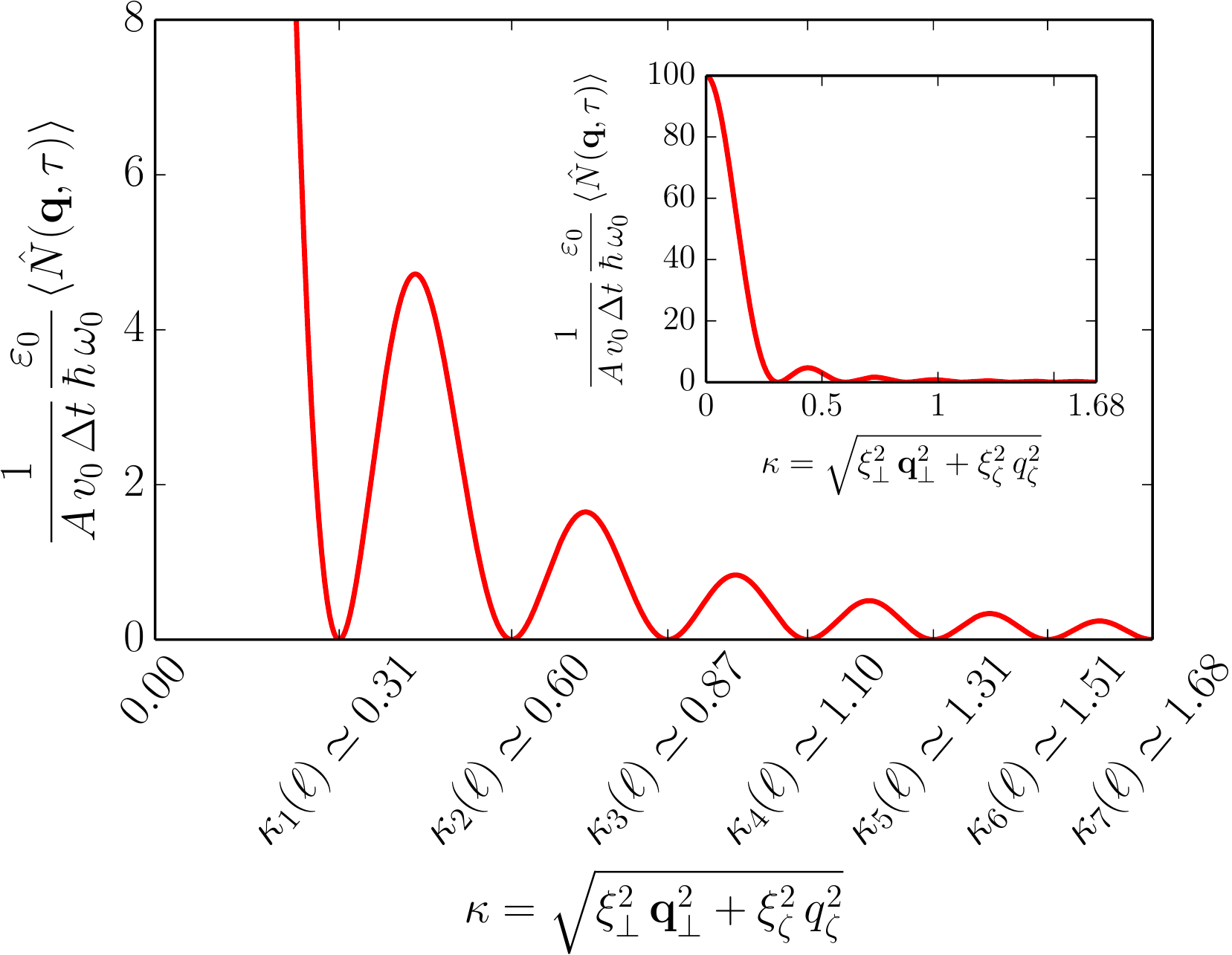}
\caption{(Color online) Normalized momentum distribution \eqref{MomentumDistribution} for $\ell=g_{0}\,\mathcal{E}^{2}\,h=10$. It reveals the occurrence of constructive and destructive interferences at well-defined periods; the $\kappa_{n}(\ell)$'s [given by Eqs.~\eqref{Periodicity}] correspond to the $\mathbf{q}$'s for which one has destructive-interference effects. The main plot focuses on the oscillations of the momentum distribution while the inset shows the whole function.}
\label{OccupationNumber}
\end{figure}

The plot of the spontaneous-four-wave-mixing intensity distribution $\langle\hat{N}(\mathbf{q},\tau)\rangle$ is shown in Fig.~\ref{OccupationNumber} for $\ell=10$. The most noticeable feature of the momentum distribution is its oscillation profile originating from $\mathbf{q}=0$: Depending on the value of the wavevector $\mathbf{q}$, the two-body emission processes at the front and the back interfaces reveal constructive or destructive interferences at well-defined periods. The latter are determined by the finite thickness $h$ of the dielectric layer and the Bogoliubov dispersion $K_{\mathrm{B}}(\mathbf{q})$ \textit{via} the product $K_{\mathrm{B}}(\mathbf{q})\,h$: For example, from Eq.~\eqref{MomentumDistribution3}, the $\mathbf{q}$'s for which one has destructive interferences, i.e., zero minima in the momentum distribution, are such that
\begin{subequations}
\label{Periodicity}
\begin{align}
\label{Periodicity1}
&\left.\frac{\ell}{2}\,\kappa\,\sqrt{\kappa^{2}+4}=n\,\pi,\right. \\
\label{Periodicity2}
\text{that is},\quad&\left.\kappa\equiv\kappa_{n}(\ell)=\sqrt{2}\;\sqrt{\sqrt{\left(\frac{n\,\pi}{\ell}\right)^{2}+1}-1},\right.
\end{align}
\end{subequations}
where $n$ is an integer $\geqslant1$.

The quantum nature of the emitted particles can be assessed \textit{via} a measurement of the two-point correlation function in the far-field regime, namely,
\begin{equation}
\label{MomentumCorrelationsDefinition}
g^{(2)}(\mathbf{q},\mathbf{q}';\tau)=\frac{\langle\hat{N}(\mathbf{q},\tau)\,\hat{N}(\mathbf{q}',\tau)\rangle}{\langle\hat{N}(\mathbf{q},\tau)\rangle\,\langle\hat{N}(\mathbf{q}',\tau)\rangle}-1,
\end{equation}
function which quantifies the correlations between the different angular and spectral components of the four-wave-mixing emission. One can evaluate \eqref{MomentumCorrelationsDefinition} by means of Wick's theorem [indeed, since the Bogoliubov-de Gennes equation \eqref{BogoliubovDeGennesEquation} is linear, the corresponding Hamiltonian---the so-called Bogoliubov-de Gennes Hamiltonian---admits a quadratic dependence on the fluctuations $\delta\hat{\mathcal{E}}$ and $\delta\hat{\mathcal{E}}^{\dag}$, and so, on the ladder operators $\hat{a}$ and $\hat{a}^{\dag}$]. Using Eq.~\eqref{InOutRelation} which connects the outgoing excitations to the ingoing ones and Eqs.~\eqref{AverageOfInterest1} and \eqref{AverageOfInterest2}, we end up finding, for all $\tau>h/(2\,v_{0})$,
\begin{equation}
\label{MomentumCorrelations}
g^{(2)}(\mathbf{q},\mathbf{q}';\tau)=\frac{(2\pi)^{3}}{A\,v_{0}\,\Delta t}\left|\frac{\tilde{u}_{\mathbf{q}}}{\tilde{v}_{\mathbf{q}}}\right|^{2}\Big[\delta(\mathbf{q}-\mathbf{q}')+\delta(\mathbf{q}+\mathbf{q}')\Big].
\end{equation}

As expected on general optical grounds and quantitatively predicted by Eq.~\eqref{MomentumCorrelations}, the two-body correlation signals in reciprocal space have to be concentrated along the (diagonal) $(\mathbf{q},\mathbf{q}'=\mathbf{q})$ and (antidiagonal) $(\mathbf{q},\mathbf{q}'=-\mathbf{q})$ lines. The first one reflects the vacuum fluctuations in the occupation $\hat{N}(\mathbf{q},\tau)$ while the second one is a clear signature of a complete correlation between excitations of opposite wavevectors $\pm\,\mathbf{q}$. The thermal analogue of this phenomenon was recently observed in a quasi-one-dimensional atomic Bose-Einstein condensate whose speed of sound was modulated \textit{via} a suitable time modulation of the transverse confinement \cite{Jaskula2012}; even though this resulted in the creation of thermal phonons of equal and opposite velocities, temperature was too high to assess the quantum origin of the observed correlations.

As a final point, it is important to highlight the novelty of our predictions with respect to standard nonlinear-optics four-wave-mixing effects, which are also described by paraxial wave equations \cite{Lin2007, Boyd2008, Zhang2009, Gao2010, Gao2011} similar \cite{Blaauboer1998, Izus1999, Blaauboer2000} to the classical counterpart of the Bogoliubov-de Gennes equation \eqref{BogoliubovDeGennesEquation}. As typical four-wave-mixing experiments benefit from maximizing the intensity of the four-wave-mixing signal, they are typically implemented in a phase-matched regime where both energy and momentum are conserved in the process: In our language, this requires that there exist $(\mathbf{k}_{\perp},\delta\omega)$ pairs for which $\mathbf{k}_{\perp}^{2}/(2\,\beta_{0})-D_{0}\,\delta\omega^{2}/2=0$, which implies that the transverse ($\mathcal{M}_{x,x}=\mathcal{M}_{y,y}=\beta_{0}$) and temporal [$\mathcal{M}_{\zeta,\zeta}=-1/(v_{0}^{2}\,D_{0}^{\phantom{2}})$] masses have opposite signs. As we have seen in Sec.~\ref{SubSec:SpatiallyHomogeneousSystem}, this is often associated to (absolute or, in some cases, only convective \cite{Izus1999}) instabilities in the quantum fluid of light.

Our angle on this physics is significantly different, as we are interested in the response of the photon fluid to quantum quenches. In an optical language, the energy that is deposed in the fluid by the quenches corresponds to the nonconservation of the normal (i.e., $z$) component of the wavevector at the $z=\mp\,h/2$ interfaces. While the four-wave-mixing intensity that results from the quantum quenches is substantially weaker than the one expected for a perfectly phase-matched configuration, it encodes a detailed information on the conservative quantum dynamics of the fluid of light in response to the sudden jumps of the interaction parameter, which is the main object of the present work. As a specific and most illuminating example of such features, the evolution of the photon fluid is ruled by the Bogoliubov dispersion law $K_{\mathrm{B}}(\mathbf{q})$ defined in Eq.~\eqref{BogoliubovSpectrum}, which involves the optical nonlinearity in a very nonperturbative way. The measurement of the spatial $(x,y)$ component of $K_{\mathrm{B}}(\mathbf{q})$, based on a bulk nonlinear medium, was recently reported in Ref.~\cite{Vocke2015}; another experiment in a one-dimensional waveguide geometry is underway to measure the temporal $\zeta=v_{0}\,t$ part of the Bogoliubov dispersion \cite{Biasi}.

\section{Conclusions and outlooks}
\label{Sec:ConclusionsAndOutlooks}

In this article, we have used a general quantum theory describing paraxial light propagation in a bulk $\chi^{(3)}$ medium in terms of quantum nonlinear Schr\"odinger evolution equations---as pioneered in Refs.~\cite{Lai1989a, Lai1989b}---to theoretically investigate the quantum statistical properties of a laser beam emerging from a finite slab of Kerr material. By mapping light propagation onto a temporal evolution and by viewing the physical time as an effective spatial coordinate, our predictions may be straightforwardly interpreted in the many-body-physics language as the response of a gas of many interacting photons to a pair of quantum quenches of the system's Hamiltonian in the nonlinear interaction parameter. This exemplifies the potential of nonlinear optics in propagating geometry as a novel platform for experimental studies of time-dependent problems in many-body physics and quantum statistical mechanics \cite{Polkovnikov2011}.

In the first sections of the present paper, we have reviewed the theory underlying the space $\longleftrightarrow$ time mapping and we have extended it to a fully three-dimensional configuration. We have then provided a comprehensive overview of its features and its power from a many-body-physics perspective. In contrast to quantum fluids of light in planar-microcavity geometries that have been considered and widely studied during the last decade \cite{Carusotto2013}, the evolution of the quantum photon field in a propagating configuration is fully conservative and not subject to the unavoidable radiative (or nonradiative) losses inherent to cavity devices: Starting from an initial many-body state that is determined by the statistical properties of the incident beam, quantum coherence can be maintained in the quantum fluid of light over macroscopic evolution times. This gives access to a \textit{plethora} of many-body dynamical quantum phenomena that can be eventually reconstructed from the statistical properties of the transmitted light using standard quantum-optics techniques.

In the second part, as a simplest, yet most remarkable example of application of the quantum formalism, we have studied in a detailed way the response of a spatially homogeneous quantum fluid of light to the pair of quenches of the interaction constant that it feels upon crossing the front and the back faces of a $\chi^{(3)}$ material immersed in vacuum. In the standard optical configuration where the incident light beam is coherent and the Kerr medium weakly nonlinear, we have demonstrated that the quantum quenches of the optical nonlinearity lead to the emission of pairs of correlated counterpropagating Bogoliubov waves on top of the laser fluid. In particular, we have pointed out how the peculiar quantum features of this process can be experimentally identified in the angular and spectral distribution as well as in the two-body correlation functions of the transmitted light in both the near and the far field.

From a general standpoint, our results illustrate the power of well-known nonlinear-optics setups as a workhorse to investigate many-body phenomena presently of great interest in quantum statistical mechanics. While we have so far focused on the configuration of easiest experimental implementation, this work paves the way to many other interesting directions: As a few examples, one can mention the study of the strong quantum fluctuations of the photon-field phase in reduced dimensions \cite{LarreFutureBis}, of the so-called Hawking radiation in analog black-hole configurations \cite{LarreFutureTer}, and, on a longer run, of the strong quantum correlations that naturally appear in the presence of significant nonlinearities at the single-photon level \cite{Lebreuilly, LebreuillyBis}.

\begin{acknowledgments}

We acknowledge Dimitris G. Angelakis, Daniele Faccio, Stefano Finazzi, Pjotrs Grisins, Christian Miniatura, and Luciano Vanzo for stimulating discussions. This work was supported by the ERC through the QGBE grant, by the EU-FET Proactive grant AQuS, Project No.~640800, and by the Autonomous Province of Trento, partly through the SiQuro project (``On Silicon Chip Quantum Optics for Quantum Computing and Secure Communications'').

\end{acknowledgments}


\begin{thebibliography}{99}
\bibitem{NoteHamiltonian}
Our choice of referring to $H(\tau)$ as a Hamiltonian follows from its interpretation as an evolution operator. However, as it generates propagation in the $z$ direction, some reader may object that it is related to the $z$ component of the linear momentum of the underlying field theory. For the sake of transparency, throughout this paper, we shall stick to the first wording.
\bibitem{NoteMomentumSpace}
To treat the infinite factor $\delta(\mathbf{q}-\mathbf{q}=0)$ appearing in Eq.~\eqref{MomentumDistribution1}, one replaces the $\delta$ function in momentum space with $\delta_{\Omega}(\mathbf{q})=\int_{\Omega}e^{i\mathbf{q}\cdot\mathbf{r}}\,d\mathbf{r}/(2\pi)^{3}$, where $\Omega$ denotes a large cylindrical volume in the $\mathbf{r}=(x,y,\zeta=v_{0}\,t)$ space whose cross section $A$ in the $(x,y)$ plane and length $L_{\zeta}=v_{0}\,\Delta t$ in the $t$ direction have the natural interpretations given in the main text. The same reasoning holds for Eq.~\eqref{MomentumCorrelations}.
\end{thebibliography}
\end{document}